\documentclass[sigconf,screen,authorversion,nonacm]{acmart}
\AtBeginDocument{%
  }

\usepackage{enumitem} % added by Max for design goal bullet styles
\usepackage{subcaption}
\usepackage{multirow}
\usepackage{makecell}

\newcommand{\system}{\textsc{Elsewise}}
\begin{document}

%\title{\system{}: Structured Exploration and Sculpting of Narrative Possibility Space with Branching Story Volume Visualization for AI-bridged Interactive Narrative}
\title{\system{}: Authoring Open-Ended Interactive Narrative with Possibility Space Visualization}

% Elsewise: Shaping Narrative Possibility Space with Structured Exploration of Emergent Branching Story Volumes for AI-bridged Interactive Narrative


\author{Yi Wang}
\email{ywang485@gmail.com}
\affiliation{%
  \institution{Independent Researcher}
  \city{San Francisco}
  \state{CA}
  \country{USA}
}

\author{John Joon Young Chung}
\email{jchung@midjourney.com}
\affiliation{%
  \institution{Midjourney}
  \city{San Francisco}
  \state{CA}
  \country{USA}
}

\author{Melissa Roemmele}
\email{melissa@roemmele.io}
\affiliation{%
  \institution{Independent Researcher}
  \city{San Francisco}
  \state{CA}
  \country{USA}
}

\author{Yuqian Sun}
\email{ysun@midjourney.com}
\affiliation{%
  \institution{Midjourney}
  \city{San Francisco}
  \state{CA}
  \country{USA}
}

\author{Tiffany Wang}
\email{twang@midjourney.com}
\affiliation{%
  \institution{Midjourney}
  \city{San Francisco}
  \state{CA}
  \country{USA}
}

\author{Shm Garanganao Almeda}
\email{shm.almeda@berkeley.edu}
\affiliation{%
  \institution{UC Berkeley}
  \city{Berkeley}
  \state{CA}
  \country{USA}
}

\author{Brett A. Halperin}
\email{bhalp@uw.edu}
\affiliation{%
  \institution{University of Washington}
  \city{Seattle}
  \state{WA}
  \country{USA}
}

\author{Yuwen Lu}
\email{ylu23@nd.edu}
\affiliation{%
  \institution{University of Notre Dame}
  \city{Notre Dame}
  \state{IN}
  \country{USA}
}

\author{Max Kreminski}
\email{mkremins@cornell.edu}
\affiliation{%
  \institution{Cornell Tech}
  \city{New York}
  \state{NY}
  \country{USA}
}

\renewcommand{\shortauthors}{Wang et al.}
\newcommand{\revision}[1]{\textcolor{black}{#1}}

%\setlength{\parskip}{2pt plus 1pt minus 1pt}

\begin{abstract}
Interactive narrative (IN) authors craft spaces of divergent narrative possibilities for players to explore, with the player's input determining which narrative possibilities they actually experience. Generative AI can enable new forms of IN by improvisationally expanding on pre-authored content in response to open-ended player input. However, this extrapolation risks widening the gap between author-envisioned and player-experienced stories, potentially limiting the strength of plot progression and the communication of the author's narrative intent. To bridge the gap, we introduce \system{}: an authoring tool for \revision{LLM-based} INs that implements a novel Bundled Storyline concept to enhance authors' perception and understanding of the narrative possibility space, allowing authors to explore similarities and differences between possible playthroughs of their IN in terms of open-ended, user-configurable narrative dimensions. A user study (n=12) shows that our approach improves author anticipation of player-experienced narrative, leading to more effective control and exploration of narrative possibility spaces.

\end{abstract}



\begin{teaserfigure}
  \includegraphics[width=0.95\textwidth]{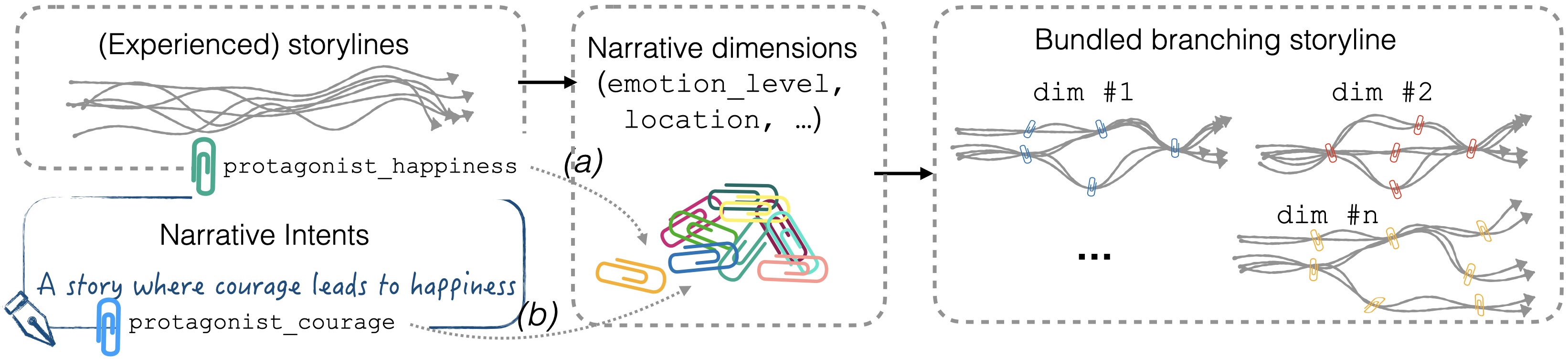}
  \centering
  \caption{We propose the novel concept of ``Bundled Storylines'' for perception and sensemaking of narrative possibility space in open-ended Interactive Narrative (IN) authoring. Given a dataset of player-experienced storylines and an author-configurable collection of open-ended narrative dimensions, we ``bundle together'' narrative states that are indistinguishable in terms of selected narrative dimensions to construct branching storyline structures. We implemented Bundled Storyline Visualization (BSV) in an LLM-based IN authoring system called \system{}. Authors can customize the visualization by creating BSVs from (a) data-derived dimensions or (b) user-defined dimensions, and creating BSVs that cross over dimensions from existing BSVs.  Leveraging a canvas of BSVs, authors can intuitively shape the distribution of narrative trajectories for their storytelling needs.}
  \Description{This figure introduces a Bundled Storyline Visualization (BSV) system designed for understanding and navigating narrative possibility spaces in AI-based interactive storytelling. The figure demonstrates how narrative states that share similar characteristics can be bundled together to create comprehensible branching structures from complex story data. Starting from experienced storylines shown as tangled gray paths and the narrative intents ("A story where courage leads to happiness"), the system extracts narrative dimensions such as emotion level and location (indicated as clips). Protagonist happiness is a dimension that is reused over again within this figure. The author themselves can also define their own dimension (protagonist courage, b). These dimensions are then used to create a bundled branching storyline representation on the right, where similar narrative paths are grouped by color-coded dimensions (also represented as clips), creating a more organized tree-like structure with dimension labels.}
  \label{bundled_branching_storyline}
\end{teaserfigure}


\maketitle

\section{Introduction}

Interactive narrative (IN) experiences permit players to explore a space of divergent narrative possibilities crafted by an author, with the player's input determining which narrative possibilities they will actually encounter
%that allows the audience to be an integral part of an authored story by meaningfully creating or influencing a storyline through their actions in the storyworld
\cite{riedl2013interactive, green2014interactive}. The most prominent form of IN features multiple branching storylines reflecting the consequences of different player actions. A fundamental challenge for IN authoring lies in the tension between two competing needs: coherently expressing the author's narrative intent ({\em authorial intent}) while immersing players by granting them meaningful agency over plot progression ({\em player agency}) \cite{riedl2013interactive}.

Conventional IN authoring workflows address this challenge by predefining a fixed range of possible player actions and crafting storylines tailored to each potential course of action---but comprehensively authoring appropriate storylines for many different potential sequences of player actions can be prohibitively burdensome, limiting the degree of player choice typically supported by conventional INs. \revision{Commercial titles such as \textit{Baldur's Gate 3}, \textit{Fallout}, and \textit{The Elder Scrolls} demonstrate that this burden can only be overcome at scale through large, well-resourced teams working over multi-year authoring, playtesting, and revision cycles.} Many authoring tools \cite{green2021use} have been introduced to help IN authors address this limitation, but the authorial burden \cite{jones2024experiencing, hargood2023authoring} associated with high-agency forms of IN remains substantial\revision{---a cost that constrains both the ambition of high-end commercial productions and the accessibility of authoring for less well-resourced creators.}

Recently, ``AI-bridged''~\cite{kim2024authors} forms of IN \revision{\cite{ai_dungeon2, inworldorigin, dramallama}} have aimed to support wide-ranging player action by using LLMs to improvise textual responses to open-ended player input. However, open-ended runtime improvisation of story content limits author insight into player experience~\cite{lu2025whatelse}; weakens plot progression as LLMs struggle to orchestrate experientially sound narratives~\cite{tian2024large, chakrabarty2024art, beguvs2024experimental}; and leads to breakdowns of expressive communication~\cite{ExpressiveCommunication} as players lose touch with the human ``hand behind the multiform plot''~\cite{HamletOnTheHolodeck}.

\revision{In response}, we propose the novel concept of {\em Bundled Storylines} to aid authors of INs in making sense of narrative possibility space (Fig.~\ref{bundled_branching_storyline}). %Our approach builds on insights from prior work in (AI-based) IN authoring \cite{lu2025whatelse, kreminski2018sketching}, storyline visualization \cite{chung2022talebrush, masson2025visual}, and general-purpose tools for visualizing and making sense of possibility spaces \cite{suh2024luminate, kreminski2022evaluating, chung2024patchview, ERA}.
A bundled storyline structure ``bundles together'' narrative states that are indistinguishable in terms of open-ended, user-configurable narrative dimensions to construct branching visualizations, which illustrate how possible playthroughs of an open-ended IN converge and diverge in an abstract narrative state space. Narrative dimensions may be defined manually by an author (Fig.~\ref{bundled_branching_storyline} (b)) to assess alignment between player experience and author's vision, or automatically derived from player-experienced storylines to reveal emergent plot dynamics (Fig.~\ref{bundled_branching_storyline} (a)). Users can also combine dimensions to cross-reference different narrative aspects.

\revision{We implement the Bundled Storyline concept in \system{}, an authoring tool for LLM-based INs---a genre that produces the most open-ended narrative possibility space and thus faces the widest gap between author-envisioned and player-experienced stories.} \system{} enables authors to define narrative possibilities in an LLM-based IN through structured storyworld initialization prompts and a trigger-based rule system adapted from storylets \cite{kreminski2018sketching}. Unlike existing authoring tools for LLM-based INs  \cite{mishra2025whatif, wu2025orchid} which focus on streamlining implementation of author's intended narrative structures, \system{} aims to assist IN authors in gradually sculpting the possibility space of player-experienced narratives. Through Bundled Storylines, authors can intuitively shape the distribution of narrative trajectories to meet their storytelling objectives. \revision{We conducted an initial author-facing user study ($n=12$), which shows} our approach effectively enables authors to better anticipate player-experienced storylines, resulting in more effective control over the narrative possibility space, while facilitating creative exploration in shaping the space. Overall, we contribute:

\begin{enumerate}
    \item The concept of {\em Bundled Storylines} to aid perception and sensemaking of player experienced storylines.
    \item \revision{\system{}\footnote{\url{https://elsewise-research.github.io}}}, an IN authoring tool that implemented Bundled Storylines and related affordances for structured exploration and sculpting of narrative possibility space.
    \item \revision{An initial author-facing} user study (n=12) that validates the effectiveness of this approach and sheds lights on how authors perceive, control and explore using our approach.
\end{enumerate}
Our work demonstrates how authoring tools can enable authors of \revision{open-ended} IN to better anticipate, explore, and shape the range of narrative outcomes that players are likely to experience.

\section{Related Work}

\subsection{(AI-Based) Interactive Narrative}

Constructing a compelling linear story already requires careful \emph{orchestration}~\cite{talmy2012narrative,Orchestration}: plot, character arcs, and low-level detail must cohere to communicate the author's expressive intent~\cite{UnmetNeeds}.
Interactivity compounds this challenge by forcing authors to contend with variation. When a story can unfold in multiple ways based on player choices, it becomes substantially harder to anticipate the event sequence any individual player will experience---and therefore to maintain control over experiential qualities such as surprise~\cite{bae2013computational}, pacing~\cite{cohn2013visual}, and tension~\cite{kybartas2020tension}. 
This difficulty of anticipating and shaping player experience forms a central part of the ``authoring problem'' discussed in IN research~\cite{AuthoringProblemIntro}.

%two of its most prominent design responses---\emph{emergent narrative}~\cite{aylett2000emergent,walsh2011emergent,ryan2015open,EmergentNarrative} and \emph{drama management}~\cite{DramaManagement,chen2009evaluating}---are often framed as different ways of maintaining authorial intent in the face of unpredictable player behavior~\cite{aylett2000emergent}.

\revision{Responses to the authoring problem vary along a broader design spectrum, defined by how much of the narrative space is fixed in advance versus generated at play-time. At one end, predefined branching storylines enumerate a discrete set of author-written paths: authorial burden grows combinatorically with the degree of player choice. Toward the middle, procedurally assembled storylets~\cite{kreminski2018sketching} decompose content into reusable units recombined via authored preconditions, trading sequencing control for a wider range of playthroughs per unit of authored content. \emph{Drama management}~\cite{DramaManagement,chen2009evaluating} is often layered on top of such structures, actively steering which units get selected at runtime to preserve authorial intent. At the far end, {\em emergent narrative}~\cite{aylett2000emergent,walsh2011emergent,ryan2015open,EmergentNarrative}  generates storylines from simulations or AI models, shifting the challenge from writing content to designing rules that keep emergent outcomes coherent with authorial intent.
}

Wherever an IN sits on this spectrum, the author is ultimately shaping a \emph{story volume}~\cite{grinblat2017emergent}: a space of possible storylines sharing common characteristics yet admitting increasing variation as narrative branching points~\cite{jones2023authorial} or recombinable content units~\cite{kreminski2018sketching} are introduced. Story volumes are \emph{closed} if they admit a finite set of possibilities---all player-experienced stories consisting of valid arrangements of a fixed set of authored content---or \emph{open} if the range of valid storylines is too wide for the author to anticipate
%in advance
~\cite{grinblat2021emergent}.
Any IN whose storylines include content not directly authored by its creator (e.g., player-written or runtime-generated text) instantiates an open story volume. \emph{AI-based}~\cite{AIBased} and \emph{AI-native}~\cite{AINative} IN, in which an AI model is invoked at runtime as part of the storytelling loop, therefore pose a particularly difficult challenge of anticipating and
shaping player experience. \revision{Our work offers a general strategy for narrowing the gap between authors' anticipation and players' experience. Though not tied to a particular type of IN, we focus our study on open-ended story volumes, since the gap is most severe.
}

\subsection{Sensemaking Tools for IN Authoring}

IN design has evolved through several increasingly complex authoring paradigms. Early tools like Twine~\cite{Twine} exemplified a \emph{branching} approach in which authors manually specify choice points that direct the storyline along discrete branches, with visualization-based tools~\cite{green2021use, green2018define} and analysis tools~\cite{partlan2018exploratory,veloso2021validating,mishra2025whatif} arising to support author understanding of the resulting structures. As authors' ambitions grew, the problem of combinatorial explosion~\cite{jones2023authorial} and the desire for greater narrative responsiveness~\cite{responsiveness} drove adoption of more modular \emph{storylet}-based structures~\cite{kreminski2018sketching}, enabling more dynamic responses to player choice but complicating authorial anticipation of all possible narrative arrangements a player might experience.

The incompatibility of storylet-based authoring with branching storyline visualization, and the need to reckon with the greater complexity of the story volumes instantiated by storylet-based IN,
%, and the need for more powerful sensemaking tools, 
encouraged new sensemaking approaches. \emph{State space} visualizations---including the \emph{Ice-Bound} authoring tools~\cite{Icebound} and \citeauthor{EmilyShortVisualizations}'s visualizations of ``narrative states''~\cite{EmilyShortVisualizations}---attempt to map the range of story states that may emerge during a playthrough.
%, indicate which content units are available at given points in the state space, and highlight coverage gaps or underserved playthrough trajectories. 
%StoryAssembler's partial branching visualizations~\cite{garbe2019storyassembler} simulate many playthroughs and reveal the implicit branching structures that emerge across subsets of them. DendryScope~\cite{otto2023dendryscope} combines playthrough simulation with static analysis to support reasoning about narrative progressions within subsets of the possibility space. 
StoryAssembler~\cite{garbe2019storyassembler} and DendryScope~\cite{otto2023dendryscope} simulate playthroughs to surface implicit branching structures and support reasoning about narrative progressions within the possibility space. 

%All of these tools attempt to help authors reckon with the greater complexity of the story volumes instantiated by storylet-based IN.

Although experiments like Fa\c{c}ade~\cite{facade} and Say Anything~\cite{SayAnything} had already introduced open-ended text input by players as a possible type of variation between storylines, it was AI Dungeon~\cite{ai_dungeon2} and similar LLM-based IN frameworks that popularized fully open-ended runtime text generation in AI-based IN~\cite{hua2020playing}. %By throwing the story volume wide open, open-text IN makes it even harder for authors to anticipate and control player experience.
So far, only a few tools have appeared to help IN authors work with runtime text generation. {\sc WhatELSE}~\cite{lu2025whatelse} distinguishes canonical (``pivot'') storylines envisioned by the author from actual player-experienced (``variant'') storylines, and allows authors to sculpt the entire story volume (or ``outline'' of the narrative space) defined by an inferred LLM prompt. Meanwhile, the Drama Llama~\cite{dramallama} and Orchid~\cite{wu2025orchid} frameworks both attempt to make LLM-based IN authoring more comprehensible by introducing storylet-based authoring affordances for shaping LLM behavior. \system{} builds directly on this line of work.

\subsection{Related Visualization Techniques}

%Outside of IN, storyline visualization has also seen uptake in tools for conventional linear narrative interpretation and construction. Story Explorer~\cite{kim2017visualizing} provided a visual toolkit for understanding temporal flow in stories told "out of order" (e.g., via flashbacks and flash-forwards). TaleBrush~\cite{chung2022talebrush} used an LLM to transform hand-drawn story curves—representing the rising and falling fortunes of a protagonist—into narrative text. Story Ribbons~\cite{yeh2025story} is an LLM-powered visual analytics workbench for analyzing how story qualities vary across a longform text, automatically extracting themes while allowing users to define their own. \citeauthor{masson2025visual}'s toolkit for \emph{visual story-writing}~\cite{masson2025visual} uses an LLM to map narrative text to structural visualizations, enabling authors both to assess their story's current structure and to revise it through direct manipulation. While all of these share a family resemblance with our approach, none supports the analysis of INs with multiple valid storylines.

Outside of IN, storyline visualization has seen uptake in tools for conventional linear narrative interpretation and construction. Story Explorer~\cite{kim2017visualizing} visualizes the flow of time in stories told ``out of order'' 
%(e.g., via flashbacks and flash-forwards)
. TaleBrush~\cite{chung2022talebrush} uses an LLM to transform hand-drawn story curves---representing a protagonist's rising and falling fortunes---into story text. Story Ribbons~\cite{yeh2025story} is an LLM-powered visual analytics workbench for analyzing how story qualities vary across a longform text, supporting both automatic and user-annotated theme extraction. \citeauthor{masson2025visual}'s \emph{visual story-writing} toolkit~\cite{masson2025visual} uses an LLM to map narrative text to structural visualizations, enabling revision through direct manipulation.
%authors to comprehend and revise their work through direct manipulation. 
However, none of these approaches support the analysis of INs with multiple storylines.

%\subsection{Possibility Space Visualization}

%\emph{Expressive range analysis} (ERA) techniques~\cite{ERA} use visualization to help creators of generative pipelines understand the distribution of outputs their systems can produce: a large number of artifacts are generated, characterized according to automatically assessable quantitative metrics, and visualized as two-dimensional heatmaps contrasting pairs of metrics. These ``slices'' through the expressive range can be summarized as a \emph{corner plot}~\cite{summerville2018expanding}, used to tune generator parameters toward designer intent~\cite{Danesh}, or combined with trajectory visualization to show how a human interactor moves through a generator's possibility space~\cite{kreminski2022evaluating,alvarez2022toward}. With the advent of LLMs, ERA-like analyses became possible along open-ended conceptual dimensions expressed as brief textual prompts rather than manually implemented evaluation functions---as in Luminate~\cite{suh2024luminate}, which organizes LLM writing suggestions along LLM-inferred conceptual dimensions, and Patchview~\cite{chung2024patchview}, which lets users browse and generate world elements on a configurable space of semantic concepts. \system{} draws on these LLM-based ERA affordances, but with a key difference: we apply ERA techniques to the space of possible \emph{narrative states} arising within an IN, rather than to the range of \emph{artifacts} produced by a procedural generator or incorporated into a creative project.

\emph{Expressive range analysis} (ERA)~\cite{ERA} uses visualization to help creators of generative pipelines (e.g., game level generators) understand the distribution of outputs their pipeline can produce. By generating a large number of artifacts and then characterizing each according to a fixed set of automatically computable metrics, the distribution of artifacts can be visualized as a set of two-dimensional heatmaps contrasting pairs of metrics. These ``slices'' through the expressive range can be exhaustively summarized as a \emph{corner plot}~\cite{summerville2018expanding}, used to tune generator parameters toward designer intent~\cite{Danesh}, or combined with trajectory visualization to show how a human interactor moves through the generator's possibility space~\cite{kreminski2022evaluating,alvarez2022toward}. 

LLMs enable ERA-like analyses over open-ended conceptual dimensions expressed as brief text labels. %, rather than manually implemented procedural evaluation functions.
Luminate~\cite{suh2024luminate} lets users generate and visually organize LLM writing suggestions along LLM-inferred conceptual dimensions; %The worldbuilding support tool
Patchview~\cite{chung2024patchview} similarly enables users to organize and generate worldbuilding elements (such as characters and settings) in a visualized configurable conceptual possibility space.
%, with affordances for configuring the conceptual space and repositioning elements. 
\system{} makes use of related LLM-based ERA affordances, but with one key difference: we apply ERA techniques to the space of possible \emph{narrative states} that may arise within an IN, rather than a space of self-contained creative artifacts. %potentially produced by a procedural generator or incorporated into a larger creative project. 

%Second, we use an LLM both to infer metrics of interest and to evaluate narrative states in terms of these metrics, rather than requiring developers to manually implement a procedural evaluation function for each metric of interest.

%[[dimensional reasoning]]

\section{Challenges of AI-Based IN Authoring}
\label{sec:design_goals}

Because storylines in open-ended INs can emerge from complex interactions between pre-authored content, player choices, and AI models, they may diverge substantially from what was envisioned during authoring.
%Unlike conventional INs, where possible storylines %experienced by the audience 
%are directly created at authoring time, storylines in AI-based INs emerge at play-time from complex interactions between the audience, the AI models, and author input, which could deviate significantly from what was envisioned during authoring. This mismatch often produces experiences that lack meaningful plot progression, leaving the author's narrative intent imperceptible to the player.
For example, in an AI Dungeon roleplay scenario, a player might begin a fantasy adventure centered on rescuing a kidnapped princess, with the author expecting the player to rescue her by confronting the villain at a designated location in town. However, if the player enters this location and defeats the villain prematurely---before the kidnapping occurs---the narrative derails from the intended storyline, leaving the player trapped outside the range of narrative posssibilities envisioned by the scenario's author and therefore unable to access the author-intended narrative content progressions that would ordinarily push the story forward in a satisfying way. To address such scenarios, authors could either guide players toward more narratively relevant actions or design additional branching storylines that accommodate certain forms of deviation while maintaining narrative coherence. However, both solutions require authors to anticipate potential player actions---a significantly more challenging task than in conventional INs, where player choices are constrained to predetermined options.

To help authors anticipate audience-experienced storylines, we distinguish between {\em intended narrative} (conceptual storylines envisioned by the author) and {\em experienced narrative} (concrete storylines experienced by the audience)~\cite{lu2025whatelse}. Intended narratives typically take the form of high-level specifications of desired narrative content, possibly conditioned on narrative state to support branching storylines. Experienced narratives are concrete narrative content experienced by players in actual playthroughs. During gameplay, the IN system synthesizes the author's specification with player input to generate narrative content (e.g., combining them into an LLM prompt). Since player input is unavailable at authoring time, the potential mismatch between the author's expectation of the player's action and the actual player's action often undermine narrative {\em control}: the intended narrative may be only partially reflected in the experienced narrative---or missed entirely.

Most existing authoring and analysis tools for open-ended INs, such as AI Dungeon~\cite{ai_dungeon2}, Inworld.ai~\cite{inworldorigin}, {\sc WhatIf}~\cite{mishra2025whatif}, Drama Llama~\cite{dramallama}, and {\sc Orchid}~\cite{wu2025orchid}, focus primarily on intended narrative, offering little to no support for understanding the experienced narratives that emerge during gameplay. Most provide only basic playtesting capabilities, resulting in highly inefficient {\em exploration} of the narrative possibility space. Meanwhile, the abstract nature of the narrative possibility space makes sensemaking inherently difficult: each storyline instance is a complex object in high-dimensional semantic space, and authors' analytical needs vary widely depending on their storytelling approaches and objectives.

% \begin{figure*}[t]
% \centering
% \includegraphics[width=\textwidth]{figures/intended_vs_experienced.png}
% \vspace{-10pt}
% \caption{Intended vs. experienced Narrative}
% ~\label{intended_vs_expereinced}
% \vspace{-10pt}
% \end{figure*}

This work aims to design authoring tools for open-ended 
%interactive narratives 
INs that provide enhanced control and exploration assistance. We approach this by %focusing on the experienced narrative and 
supporting systematic perception and configurable sensemaking of the possibility space of {\em experienced} narratives. We propose the following design goals for such authoring tools:

\begin{enumerate}[label=DG\arabic*.]
    \item {\bf Player Perspective.} The tool should incorporate player perspectives into the authoring process to surface potential discrepancies between intended and experienced narrative, helping authors align the two.
    \item {\bf Comprehensiveness.} %To be more comprehensive than simple playtesting, 
    The system should assist authors to {\em systematically} perceive the possibilities residing in the space of experienced narrative, in helping the author better anticipate the possible storylines a player could experience.
    \item {\bf Configurability.} The system should provide customizable affordances for sensemaking of the experienced narrative possibility space, tailored to the author's specific storytelling goals and story content.
\end{enumerate}

% \begin{itemize}
%     \item {\bf DG1. Assist authors to better perceive the possibility space of experienced narrative.} The system should help the author to systematically explore the narrative space, and better anticipate possible storylines a player could experience.
%     \item {\bf DG2. Assist authors to better control the narrative possibility space.} Authors control narrative space through their specification of the intended narrative. Due to the mismatched expectations on player actions, the author's change in their specification may not be significantly reflected in player's experience. The system should help the author to more effectively impact player's experience through changes in their specification.
%     \item {\bf DG3. Assist authors to better express their narrative intents.} The system should help the authors to feel more expressive through the anticipated player experience by making them more confident that their narrative intents are effectively communicated.
% \end{itemize}

\section{Bundled Storylines}

\subsection{Bundled Storyline Concept}

% \begin{figure*}[t]
% \centering
% \includegraphics[width=\textwidth]{figures/bundled_branching_storylines-v1.png}
% \vspace{-10pt}
% \caption{Bundled Branching Storyline Visualization}
% ~\label{bundled_branching_storyline}
% \vspace{-10pt}
% \end{figure*}

Can the open-ended story volumes defined by the interplay between author intent, player choice, and AI improvisation in open-ended IN be made as readily comprehensible to authors as the intuitive flowchart-like branching structures of conventional IN? 
%Although explicitly defined branching storylines are no longer the underlying mechanism of narrative generation in AI-based IN systems, a branching visual representation remains a familiar and powerful sensemaking tool.
We introduce the concept of {\em Bundled Storylines} %to capture both the emergent nature of experienced narrative and the internal dynamics of the storyworld, helping authors perceive and navigate the narrative possibility space.
to bridge the gap between these representations of IN and help authors of open-ended IN perceive and sculpt the narrative possibility space.

An experienced storyline is a sequence of events (``fabula'' instead of ``sjuzhet'' \cite{bal2009narratology}) dynamically constructed through interactions between the player and non-player characters, describing how the storyworld changes over time \cite{riedl2010narrative}. Different storylines can thus be viewed as  distinct trajectories in the possibility space 
%$S$
of all possible states of the storyworld.

\revision{We thus view an {\em experienced storyline} as a sequence of narrative states unfolding over a shared timeline, so that states from different storylines at the same timestep are directly comparable. Given a set of narrative dimensions capturing relevant plot conditions (e.g., \textit{tension level}, \textit{friendship between two characters}, \textit{whether Aladdin has the magic lamp}, etc.), we can position each narrative state in a possibility space defined by these dimensions, allowing storylines to be comprehensively compared as trajectories through that space.}

%We thus view an {\em experienced storyline} $T_i$ as an ordered sequence of {\em narrative states} $\langle s^i_1, s^i_2, \dots, s^i_{{\bf t}_{i}}\rangle$ $({\bf t}_i > 0)$. Given a set ${\bf T}$ of experienced storylines, we assume a shared global timeline, so that $s^i_t$ and $s^j_t$ (for any $i \neq j$) are comparable as different narrative states at the same timestep.  We use $S_{\bf T}$ to denote the set of all narrative states across ${\bf T}$. Given an arbitrary set of {\em narrative dimensions} representing relevant plot conditions (e.g., \textit{tension level}, \textit{friendship between two characters}, \textit{whether Aladdin has the magic lamp}, etc.), a possibility space of narrative states can be constructed and used to position each narrative state relative to others based on their value on these dimensions, allowing the experienced storylines (${\bf T}$) to be comprehensively compared as trajectories in this space. 

%An experienced storyline can be represented in different forms---a script, a dialog/action history, a sequence of images, or a video. We assume any such representation can be split into a sequence of fragments $\langle o^i_1, o^i_2, \dots, o^i_{{\bf t}_{i}}\rangle$ so that each $o^i_t$ describes the latent underlying narrative state $s^i_t$. We denote the set of all narrative space representations as $O$. For a narrative dimension $d$ with value range $dom(d)$, we assume a function $f_d(\cdot): O \rightarrow dom(d)$ exists to extract the value of $d$ from any narrative state representation.  

\revision{An experienced storyline can be represented in different forms---a script, a dialog/action history, a sequence of images, or a video. We assume any such representation can be split into a sequence of fragments $\langle o^i_1, o^i_2, \dots, o^i_{\bf T}\rangle$ so that each $o^i_t$ {\em describes} the latent underlying narrative state at timestep $t$ for storyline $i$. We denote the set of all narrative space representations as $O$. For a narrative dimension $d$ with value range $dom(d)$, we assume a function $f_d(\cdot): O \rightarrow dom(d)$ exists to extract the value of $d$ from any narrative state representation.  }

Although the notion of bundled storylines is in principle generalizable to other story representations, this paper focuses on textual representation, where each $o^i_t$ is a text fragment and the dimension extraction functions $f_d(\cdot)$ can be implemented via language models and parameterized prompts. Since we work with textual representations rather than the underlying narrative states directly, we use the term ``narrative state'' hereafter to refer to these text representations. We also restrict the focus of this work to {\em finite}, {\em discrete} value ranges for all narrative dimensions.

\begin{example}
\label{eg:storylines}
Consider the storyworld described in Table~\ref{tab:example_story_domain}, three experienced storylines in Table~\ref{tab:example_storylines}, and two narrative dimension $d_1$ (``ducks' advantage against goose'')
%($dom(d_1)=\{low, medium, high \}$)
, and $d_2$ ( ``duckling behavior'')
%($dom(d_2)=\{passive, neutral, proactive \}$).
Table~\ref{tab:example_trajectories} shows example narrative dimension values extracted from the three storylines at each timestep.\footnote{Note that the values of narrative dimension extracted from different storylines can differ based on specific LLM model and prompt used.}
\vspace{-3mm}
\end{example}

\begin{table}[b]
\footnotesize
\vspace{-10pt}
% \resizebox{\linewidth}{!}{
\caption{Example Storyworld}
\vspace{-10pt}
\begin{tabular}{p{1.1cm} | p{7.1cm}}
\hline
{\bf World Setup} & The ducks and geese sharing the pond grow increasingly agitated as winter approaches. With food becoming scarce, competition intensifies among the waterfowl, filling the air with tension and the sharp sounds of territorial disputes. \\\hline
{\bf Characters} & {\bf Duckling} (Protagonist) - A weak duckling who's just a kid and do not know much about the world.

\vspace{0.5mm}

{\bf Duck Mom} - Duckling's Mom. She loves duckling. She's torn apart between teaching duckling the cruel reality of the world and give him a happy and worry-free childhood.

\vspace{0.5mm}

{\bf Goose} - The bully in the pond. Very strong and territorial. Mean creature.
\\\hline
\end{tabular}
\vspace{2pt}
\Description{A two-column table titled "Example Storyworld" describing a narrative scenario. The first column lists structural elements and the second column provides details. The World Setup describes ducks and geese on a shared pond facing increasing conflict as winter brings food scarcity. The Characters section introduces three figures: the Duckling, a naive young protagonist; Duck Mom, who is conflicted between sheltering her child and teaching him harsh realities; and Goose, an aggressive and territorial bully.}
\label{tab:example_story_domain}
\end{table}

\begin{table}[t]
\footnotesize
% \resizebox{\linewidth}{!}{
\caption{Example Storylines for the storyworld described in Table\ref{tab:example_story_domain} ($t$ denotes timestep)}
\vspace{-7pt}
\begin{tabular}{p{0.1cm} | p{2.7cm} | p{2.1cm} | p{2.3cm}}
\hline
{\bf $t$} & {\bf Storyline 1 ($o^1_t$)} & {\bf Storyline 2 ($o^2_t$)} & {\bf Storyline 3 ($o^3_t$)} \\ \hline
1 & The duck mom warns the duckling that winter scarcity is making the geese territorial and aggressive. & The duck mom defends the duckling against the threatening goose. & The duck mom confronts the goose defending her duckling from mockery. \\ \hline
2 & The goose issues an ultimatum to exile the ducks by sunset, refusing their peaceful negotiations. & The duckling rallies other ducks to confront the goose as a group.& The duckling chases ripples, meeting an aggressive goose. \\ \hline
3 & The ducks are forced to surrender their territory and leave. &  United, the ducks successfully protect their territory from the goose.
 & The duck mom protects the curious duckling from the goose defending its territory.\\ \hline
\end{tabular}
%\vspace{2pt}
\Description{A table showing three example storylines across three timesteps for a storyworld involving ducks and geese. Storyline 1: at timestep 1, the duck mom warns the duckling that winter scarcity is making the geese territorial and aggressive; at timestep 2, the goose issues an ultimatum to exile the ducks by sunset, refusing their peaceful negotiations; at timestep 3, the ducks are forced to surrender their territory and leave. Storyline 2: at timestep 1, the duck mom defends the duckling against the threatening goose; at timestep 2, the duckling rallies other ducks to confront the goose as a group; at timestep 3, united, the ducks successfully protect their territory from the goose. Storyline 3: at timestep 1, the duck mom confronts the goose defending her duckling from mockery; at timestep 2, the duckling chases ripples, meeting an aggressive goose; at timestep 3, the duck mom protects the curious duckling from the goose defending its territory.}
\label{tab:example_storylines}
 \vspace{-10pt}
\end{table}

\begin{table}[t]
\footnotesize
% \resizebox{\linewidth}{!}{
\caption{Temporal progression of narrative dimension values for the example storylines described in Table~\ref{tab:example_storylines} ($t$ - timestep)}
\vspace{-7pt}
\begin{tabular}{p{0.2cm} | p{1cm} | p{1cm} | p{1cm} | p{1cm} | p{1cm} | p{1cm} }
\hline
{\bf $t$} & {\bf $f_{d_1}(o^1_t)$} & {\bf $f_{d_2}(o^1_t)$} & {\bf $f_{d_1}(o^2_t)$} & {\bf $f_{d_2}(o^2_t)$} & {\bf $f_{d_1} (o^3_t)$} & {\bf $f_{d_2}(o^3_t)$}\\ \hline
1 & medium & neutral & low & passive & medium & neutral\\ \hline
2 & low & passive & medium & proactive & medium & proactive\\ \hline
3 & low & passive & high & proactive & medium & neutral\\ \hline
\end{tabular}
\vspace{2pt}
\Description{Table showing the temporal progression of narrative dimension values across three timesteps for three example storylines. Columns are: timestep (t), and for each of three storylines (o1, o2, o3), values for dimension d1 and dimension d2. At timestep 1: storyline o1 has d1=medium, d2=neutral; storyline o2 has d1=low, d2=passive; storyline o3 has d1=medium, d2=neutral. At timestep 2: storyline o1 has d1=low, d2=passive; storyline o2 has d1=medium, d2=proactive; storyline o3 has d1=medium, d2=proactive. At timestep 3: storyline o1 has d1=low, d2=passive; storyline o2 has d1=high, d2=proactive; storyline o3 has d1=medium, d2=neutral.}
\label{tab:example_trajectories}
\vspace{-10pt}
\end{table}

\subsubsection{Bundled Storyline as a Timeline of  Single Narrative Dimension (``1D BSVs'')}

Given a single narrative dimension $d$, we construct a graph $G_d$ representing a branching storyline structure as follows: 
\begin{itemize}
\item {\bf Narrative states sharing the same value of $d$ at the same time are grouped together as a node.}\hspace{1mm} \revision{For each $v\in dom(d)$ and timestep $t$, $G_d$ has a node $O^{v}_t$ containing every narrative state $o^i_t$ with $f_d(o^i_t) = v$ (omitted if empty);}
\item {\bf Edges are possible ``1-step'' transitions between different values.}\hspace{1mm} \revision{An edge runs from node $O^{v_1}_{t}$ to $O^{v_2}_{t+1}$ whenever some storyline $i$ satisfies $o^i_t\in O^{v_1}_{t}$ and $o^i_{t+1}\in O^{v_2}_{t+1}$.}

\end{itemize}
Intuitively, we ``bundle'' together all the narrative states at the same timestep sharing the same value in dimension $d$ to construct the nodes in the graph, and merging temporal connections sharing both starting and ending bundles. This gives us a {\em Bundled Storylines} graph of the experienced storylines relative to the dimension $d$. 

\begin{figure}[tbp]
\centering
\includegraphics[width=0.48\textwidth]{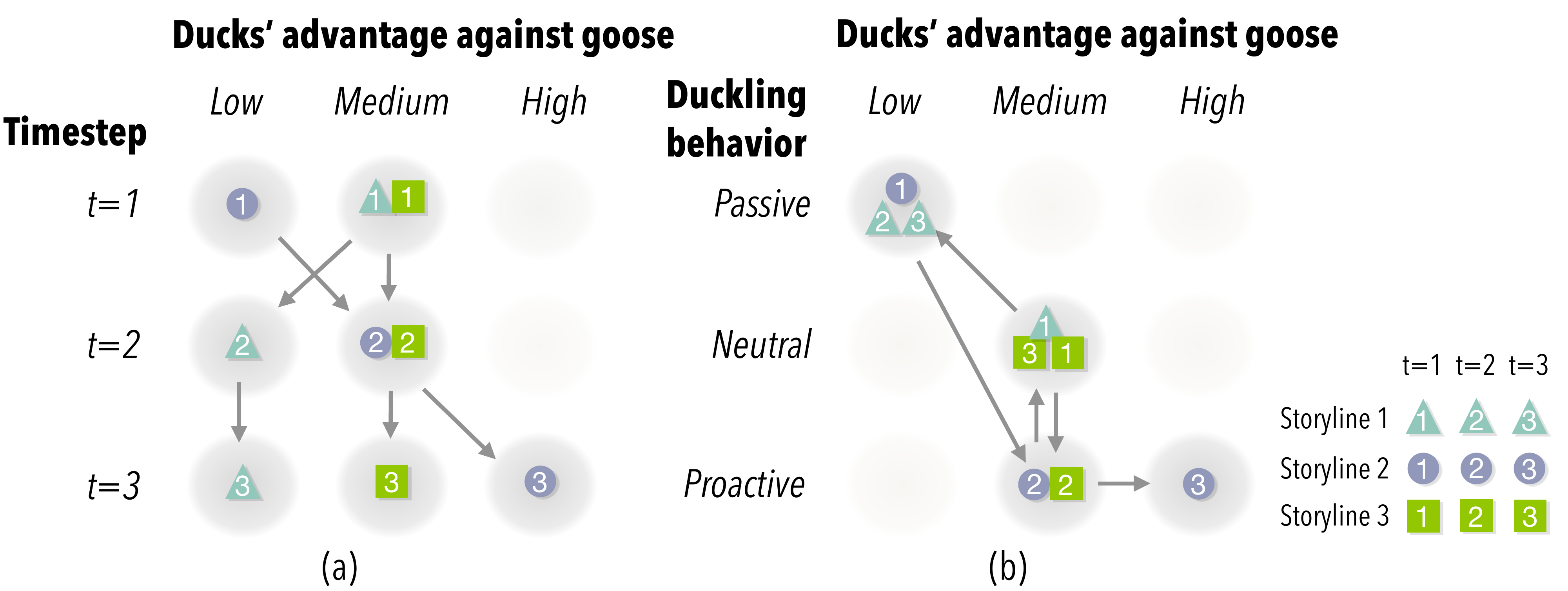}
\caption{Example Bundled Storylines for single narrative dimension (a) and two narrrative dimensions (b)}
\Description{The figure contains two related state-transition diagrams, labeled (a) and (b), both sharing the x-axis label "Ducks' advantage against goose" with three levels: Low, Medium, and High.
Diagram (a) illustrates state transitions across three timesteps (t=1, t=2, t=3). Each state is represented as a gray bubble containing one or more colored dots symbolizing narrative states (with blue dots, teal dots and green dots representing three distinct storylines). At t=1, there is one bubble in the Low column containing a single blue dot, and one bubble in the Medium column containing a teal and a green dot. Arrows from t=1 cross between columns: the Low-column bubble transitions to the Medium column at t=2, and vice versa (crossed arrows). At t=2, the Low column holds a teal dot, and the Medium column holds a blue and green dot. From t=2, a downward arrow keeps the Low-column bubble in the Low column at t=3 (now showing a teal dot), while the Medium-column bubble splits: one arrow goes straight down to a green dot in the Medium column, and a diagonal arrow goes right to a blue dot in the High column at t=3.
Diagram (b) adds a second dimension: "Duckling behavior" on the y-axis with three levels: Passive (top), Neutral (middle), and Proactive (bottom). States are again represented as gray bubbles with colored dots. In the Low column at the Passive level, a bubble contains a blue and teal dot. In the Medium column at the Neutral level, a bubble contains a large green dot and a small teal dot. In the Low column at the Proactive level, a bubble contains a blue and green dot, and in the High column at the Proactive level, a bubble contains a single blue dot. Arrows show: the Passive/Low bubble pointing down-right to the Neutral/Medium bubble; the Neutral/Medium bubble pointing down-left to the Proactive/Low bubble; a bidirectional arrow between Proactive/Low and Neutral/Medium; and a rightward arrow from Proactive/Low to Proactive/High.}
~\label{example_bsv}
\vspace{-20pt}
\end{figure}

\revision{In our implementation, we use dots of different colors to represent narrative states from different storylines. For each bundled storylines graph, we arrange dots in a matrix whose columns correspond to values of the narrative dimension, and rows correspond to each timestep
%so that dots representing narrative states are positioned at the row representing the corresponding timestep and the column representing corresponding value of the dimension  
(Fig.~\ref{example_bsv} (a)). We call this representation of a bundled storylines graph a {\em Bundled Storyline Visualization}, or a (1D) BSV.}

\begin{example}
\vspace{-2mm}
%Consider the narrative dimension $d_1$ from Example~\ref{eg:storylines}. 
The BSV of $G_{d_1}$ (Figure~\ref{example_bsv} (a)) shows that ducks' advantage over geese typically starts low or medium and follows one of three trajectories: remaining constant (undesirable lack of progression), declining to "low" (unsuccessful playthrough), or rising to "high" (successful playthrough).
\vspace{-2mm}
\end{example}

\subsubsection{Bundled Storyline with Multiple Narrative Dimensions (``2D BSVs'')}\label{sec:2d_bsv}
It is theoretically possible to extend the definition of bundled storyline to multiple narrative dimensions. Given a subset of dimensions $D = \{d_1, \dots, d_n\}$, the graph $G_D$ representing a branching storyline structure is constructed as follows:

\begin{itemize}
\item {\bf Narrative states that share the same value combination at the same timestep are grouped together as a node.}\hspace{1mm} \revision{For each distinct value combination ${\bf v} = v_1, \dots, v_n$ ($v_1\in dom(d_1), \dots, v_n\in dom(d_n)$) and timestep $t$, $G_D$ has a node $O^{\bf v}_t$ containing every narrative states $o^i_t$ with $f_{d_k}(o^i_t) = v_k$ for all $k$ (omitted if empty);}
\item {\bf Edges represent possible ``1-step'' transitions between different value combinations.}\hspace{1mm} \revision{An edge runs from node $O^{{\bf v}_1}_{t}$ to $O^{{\bf v}_2}_{t+1}$ whenever some storyline $i$ satisfies $o^i_t\in O^{{\bf v}_1}_{t}$ and $o^i_{t+1}\in O^{{\bf v}_2}_{t+1}$.}
\end{itemize}

Intuitively, narrative states sharing the same value combination on the dimensions are indistinguishable in this subspace, forming a ``bundle'' of narrative states. We call $G_D$ a Bundled Storyline graph of the experienced storylines relative to the dimension set $D$.

A {\em two-dimensional Bundled Storyline Visualization} arranges dots representing narrative states in a matrix whose axes correspond to the possible values of two narrative dimensions (Fig.~\ref{example_bsv}(b)). Dots sharing the same dimension-value combination occupy the same matrix cell, regardless of variation along other aspects. We cap the number of narrative dimensions at two to avoid the visual complexity of higher-dimensional matrices.
Conceptually, an additional dimension representing timestep can be added to the matrix to incorporate temporal information. To preserve a 2D visualization, we adopt the time flattening technique~\cite{bach2014review}, encoding temporal order through dot size (smaller dots correspond to earlier narrative states). Admittedly, it is difficult to visually distinguish differences in sizes to quickly get precise temporal relations, but the focus of 2D BSV views is to reveal correlations across narrative dimensions rather than track the progression of any single one. Methods for incorporating temporal information when cross-referencing multiple narrative dimensions are described in Section~\ref{sec:highlighting_mechanism}.
\vspace{-1mm}
\begin{example}
%Consider the narrative dimensions $d_1$ and $d_2$ from Example~\ref{eg:storylines}.
Figure~\ref{example_bsv} (b) shows the BSV of $G_{\{d_1, d_2\}}$. This graph focuses on the correlation between ducks' advantage against goose and duckling behavior, rather than the temporal progression of any single narrative dimension. Assuming the storylines are representative, the data suggest that proactive duckling behavior tends to correlate with greater advantage against goose, while passive behavior correlates with diminished advantage.
\end{example}
\vspace{-1mm}
Like branching storylines for conventional IN, the bundled storyline representation  captures possible transitions between value combinations of the selected dimensions in a state machine-like structure (DG2). The structure resembles branching storylines when abstracting away each graph node's internal structure---but with a key difference: by dynamically deriving nodes from arbitrary storylines and narrative dimensions, branches emerge from actual playthroughs rather than being predefined by the IN author (DG1).

%For INs with open-ended generative narrative content, an explicit schema of such dimensions is often missing. On the other hand, LLMs allow us to extract semantic feature from any arbitrary aspect for any narrative state. We thus propose to define a narrative state as a collection of infinite amount of semantic features, and support posterior creation of customized dimensions to extract relevant semantic features from a given set of experienced storylines. We support two interactions to create such dimensions:

\subsection{Bundled Storyline Affordances}
\label{sec:interactions}

%The previous section defines Bundled Storylines. 
This section explains the origins of narrative dimensions and how they collectively supports perception and sensemaking of the narrative possibility space. We design affordances of Bundled Storylines in support of configurable sensemaking (DG2, DG3), organized in two categories: {\em dimension extraction} and {\em dimension crossreferencing}.

\subsubsection{Dimension Extraction}\label{sec:dimension_extraction}

Bundled Storylines bridge authors' narrative vision and players' narrative experience by helping authors understand emergent plots, identify misalignments between their intent and audience experience, and refine the narrative space to steer storylines toward meaningful directions. To this end, narrative dimensions should draw from both the author's vision and players' experience, with relative emphasis determined by the authoring approach and analytical objectives. We provide three dimension extraction mechanisms that enable configurable weighting of author vision versus player experience (DG3).

{\bf Author vision only}\hspace{1mm} The user can add a customized dimension by %entering a name for the dimension (e.g., ``\textit{duckling\_courage}''), a line of text 
describing what it is about (e.g. ``\textit{how courageous the duckling feels}''), and its values (e.g., ``\textit{not\_courageous}, \textit{partially}, \textit{courageous}'').

{\bf Player experience only}\hspace{1mm} In this case, both the narrative dimension and its values are generated automatically by analyzing the dataset of player experienced storylines through {\em concept induction}~\cite{lam2024concept}, a computational process that produces interpretable high-level concepts %(possibly defined by explicit inclusion criteria)
from unstructured text (narrative state representations). %In summary, the system 1)  analyzes the playthrough data and extracts a set number of narrative dimensions that characterizes how different narrative states differ from each other; 2) analyzes the given set of experienced storylines in terms of the new dimension; and 3) extracts categorical values for this dimension by clustering the narrative states from the experienced storylines. 
Since dimensions created in this way are fully derived from the given dataset of player experienced storylines, we also call them ``data-derived dimensions'' in the rest of the paper.

{\bf Mixed}\hspace{1mm} The narrative dimension and values are collaboratively defined by both the author and player-experienced storylines. For example, the author may name and describe the dimension while the system extracts categorical values via concept induction---or conversely, the system extracts narrative dimensions by analyzing narrative states while the user defines the categorical values.

%Regardless of how dimensions are extracted, 
Once a dimension and its values are established, the system uses an LLM to classify narrative states against those values as labels, constructing the BSV view for that dimension. 
%When values are generated from experienced storylines, the clustering mechanism ensures that every value applies to at least one narrative state. This is not guaranteed for user-defined values — for instance, the duckling may never exhibit courage across all playthrough data, leaving the \texttt{courageous} category empty. Even so, such values remain useful when an author wants to refine the game to steer the narrative in that direction.

\subsubsection{Dimension Cross-referencing}\label{sec:dimension_cross_referencing}

The user gets a holistic understanding of the narrative possibility space not from a single BSV, but a collection of BSVs presenting information from different aspects (called a ``BSV canvas''). We design the following mechanisms for the user to cross-reference between multiple BSVs. 

{\bf Combining Two 1D BSVs into a 2D BSV}\hspace{1mm} 
\label{sec:cross_dimensions} To closely examine the correlation between two narrative dimensions, the user can combine dimensions from existing 1D BSVs to create 2D BSVs, facilitating sensemaking of the experienced storylines by cross-referencing two aspects of the narrative state space (DG2, DG3). For example, combining the timeline view of ``\texttt{duck's advantage against goose}'' and the timeline view of ``\texttt{duckling behavior}'' will result in the 2D BSV in Fig.~\ref{example_bsv} (b).

{\bf Cross-referencing Dimensions across Multiple BSVs}
\label{sec:highlighting_mechanism}\hspace{1mm} 
Although a single BSV is limited to no more than two narrative dimensions, correlations beyond two dimensions can be examined by cross-referencing multiple BSVs (DG2, DG3). To support this, we introduce two narrative state highlighting mechanisms:
\begin{enumerate}
\item {\em filtering by storyline} (Fig. ~\ref{multiple_bsv_with_storyline_filter}, Appendix): highlighting {\em all the narrative states from the same storyline} in all BSVs for tracking temporal progression from all explored dimensions. 
\item {\em filtering by dimension values} (Fig. \ref{multiple_bsv_with_dimension_value_filter}, Appendix): highlighting {\em all the narrative states with this value (combination)} in all BSVs to indicate potential correlation %between this valuation 
to other dimensions. 
\end{enumerate}

{\bf Timeline Control Slider (Fig. \ref{multiple_bsv_with_timeline_slider}, Appendix)}\hspace{1mm} \label{sec:timeline_control_slider} To support temporal analysis across multiple BSVs simultaneously, we provide a timeline slider that allows users to select a specific timestep of interest. The corresponding narrative states are then highlighted across all BSVs on the canvas, enabling users to trace the temporal progression of multiple narrative dimensions in parallel. 

%To summarize, 
Table~\ref{tab:sensemaking_types} lists specific types of sensemaking a BSV canvas supports.
%through dimension cross-referencing.

\begin{table}[tp]
\footnotesize
% \resizebox{\linewidth}{!}{
\caption{Types of sensemaking a BSV canvas supports}
\vspace{-5pt}
\begin{tabular}{p{4cm}|p{4cm}}
{\bf Type of Sensemaking} & {\bf Cross-referencing Mechanism} \\\hline\hline
Temporal progression of single narrative dimension & 1D BSVs (Fig~\ref{example_bsv} (a)) \\ \hline
Correlation between two narrative dimensions& 2D BSVs (Fig~\ref{example_bsv} (b)) \\ \hline
Correlation between more than two narrative dimensions & Multiple BSVs with dimension value filter and storyline filter (Fig~\ref{multiple_bsv_with_dimension_value_filter}, Fig~\ref{multiple_bsv_with_storyline_filter}, Appendix) \\ \hline
Temporal progression of correlation between multiple narrative dimensions & Multiple BSVs with timeline control slider (Fig~\ref{multiple_bsv_with_timeline_slider}, Appendix) \\ \hline
\end{tabular}
\Description{A two-column table with four rows of content. The columns are headed "Type of Sensemaking" and "Cross-referencing Mechanism." The first row describes the temporal progression of a single narrative dimension, supported by one-dimensional BSVs. The second row describes the correlation between two narrative dimensions, supported by two-dimensional BSVs. The third row describes the correlation between more than two narrative dimensions, supported by multiple BSVs with a dimension value filter and a storyline filter. The fourth row describes the temporal progression of correlation between multiple narrative dimensions, supported by multiple BSVs with a timeline control slider.}
\label{tab:sensemaking_types}
\vspace{-10pt}
\end{table}

\section{\system{}: Authoring with BSV}

To demonstrate Bundled Storyline's usefulness for authoring open-ended INs, we present IN authoring system \system{}, describing its interface, authoring affordances and their interplay with BSV.

\begin{figure*}[t]
\centering
\includegraphics[width=0.95\textwidth]{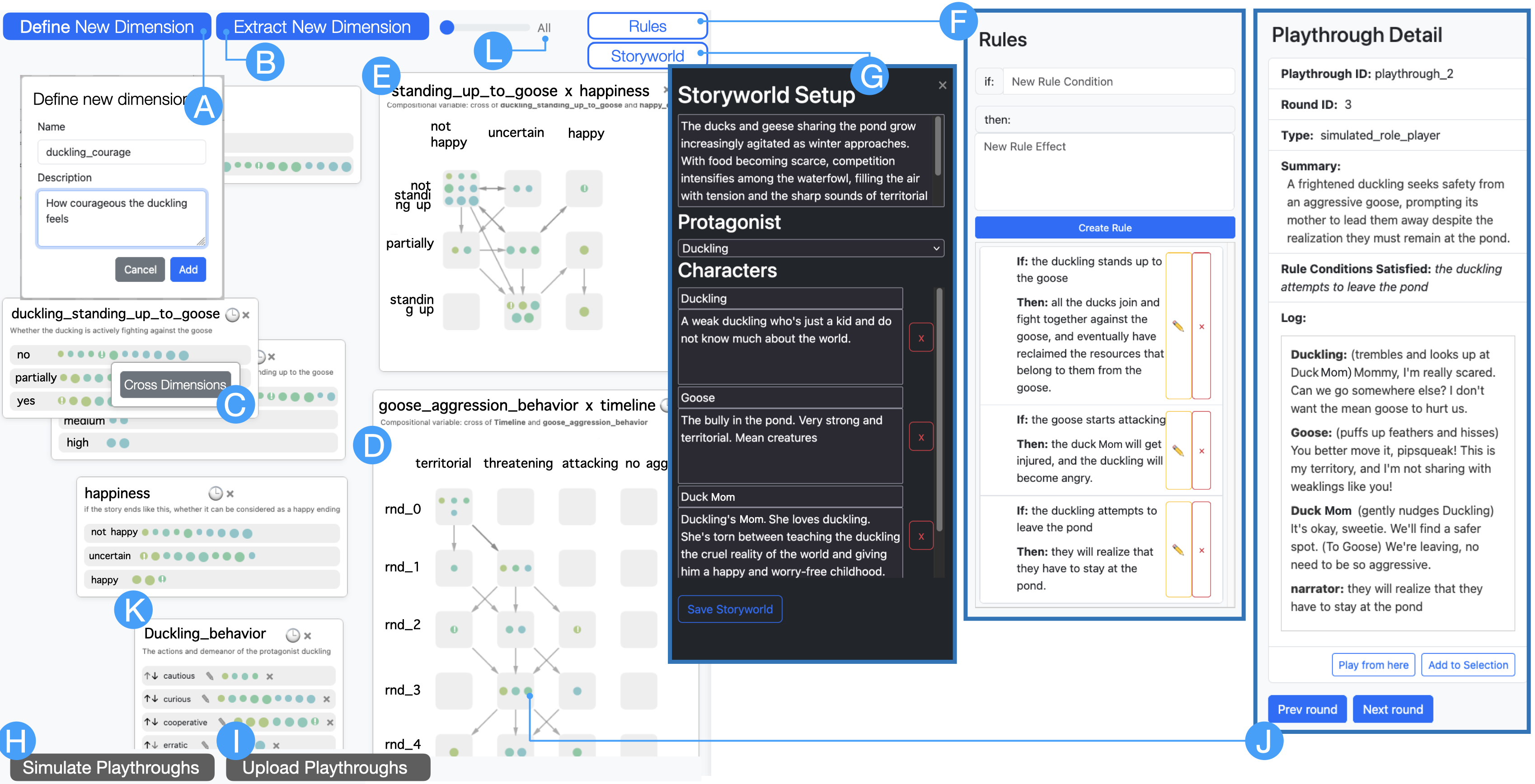}
\vspace{-10pt}
\caption{Illustration of the \system{} user interface (font enlarged for readability). Dots represent narrative states in experienced storylines. Dots of the same color are from the same storyline. An exclamation mark on a dot indicates that a rule was triggered at the corresponding narrative state. (A) Users can create BSVs by defining new narrative dimensions. (B) Users can create BSVs by prompting the system to extract a set number of data-derived narrative dimensions with concept induction on playthrough data. (C) Users can create 2D BSVs by crossing dimensions from existing BSVs. (D) A 1D BSV with timeline. (E) A 2D BSV with row and column each representing a different dimension. (F) Rule editing interface. (G) Storyworld editing interface. (H) Users can prompt the system to generate a set of simulated playthrough. (I) Users can upload playthrough data to the system. (J) When users click on a dot representing a narrative state, a panel shows up to provide details of the narrative state, and allows the user to quickly navigate to the previous and next narrative state on the same storyline. (K) Compact 1d BSV views which simply classifies all the narrative states into each categorical value of the dimension. (L) Timeline control slider. }
\Description{This figure demonstrates the Elsewise interface for creating and manipulating Bundled Storyline Visualizations in AI-based interactive narrative authoring. The interface shows multiple panels working together to help authors understand and shape narrative possibility spaces. Panel A shows the dimension definition interface where users create new narrative dimensions. The example shows "duckling_courage" being defined with a description about how courage relates to the duckling character's feelings.  Panel B presents the dimension extraction button, allowing users to prompt the system to automatically extract meaningful narrative dimensions from playthrough data.  Panel C displays a cross-dimension interface where users can create higher-dimensional BSVs by combining existing dimensions. Penal D shows an example of timeline view of an one-dimensional BSV "goose_aggression_behavior × timeline" with timeline values from rnd_0 through rnd_4 representing different story rounds. Panel K illustrates two one-dimensional BSVs--- the first one (top) for "happiness," showing narrative states classified into categories from "not happy" through "uncertain" to "happy,"; and the second one (bottom) "Duckling_behaviour" showing narrative states classified into categories from "cautious", "curious", "cooperative" and "erratic", with editting mode enabled to add/delete/change values of the narrative dimension. Panel E demonstrates a two-dimensional BSV with a grid where rows and columns each represent different dimension values, standing_up_to_goose and happiness, allowing for more complex narrative state organization. Panels F and G show the rule and storyworld editing interfaces. The rule editor (F) allows authors to define conditions and effects that govern story progression. The storyworld editor (G) provides fields for setting up characters, locations, and narrative elements like "Protagonist: Duckling" and various character relationships. Panel H simulates playthroughs button, while Panel I shows a button to upload existing playthrough data. Panel J shows the playthrough detail view that appears when users click on narrative state dots, displaying the specific story content and allowing navigation to previous and next states along the same storyline.}
~\label{system_interface}
\vspace{-15pt}
\end{figure*}

\subsection{\system{} Interactive Stories}

The open-ended IN that authors create with \system{} uses an LLM to drive conversational gameplay according to author-defined narrative specifications. During gameplay, the LLM acts as a game master (as in tabletop RPGs like \emph{Dungeons \& Dragons}~\cite{wizards2014player}), serving as a proxy for the author. Drawing on the most prominent forms of author input found in existing AI-based IN systems~\cite{ai_dungeon2, inworldorigin, dramallama, wang2024storyverse, lu2025whatelse}, we distill the author specification into two core parts:

\begin{itemize}
    \item A {\bf storyworld} description serving as long-term context, comprising a world description and a set of characters (each described in natural language), with one character marked as the protagonist to be role-played by the player.
    \item A list of {\bf rules} governing the LLM game master's behavior, each pairing a {\em condition} with an {\em effect} in natural language. Acting as the author's proxy, the game master steers the narrative toward the effect whenever the condition is met.
\end{itemize}

An instance of the above specification defines a playable interactive story. During gameplay, the LLM game master and player alternate turns: the game master generates narration, non-protagonist dialogue, and actions, while the player controls only the protagonist. A narrative state in this context is defined as a single round of play (one turn each) encompassing the conversation history to that point. At the end of each round, the system evaluates all rules against the current narrative state and executes any whose conditions are met.
%(See Fig. \ref{} for main game loop design and Appendix \ref{} for LLM prompts used at each stage).

%Our design of the AI-based IN game mechanism captures most representative elements from existing AI-based IN systems. The two prominent types of authorial input are 

%The next section describes the authoring interface assisting the author in crafting the interactive story specifications.

\subsection{IN Authoring Interface}

\label{sec:interactive_stories}
\system{} implements BSV with a {\bf narrative space visualizer}, and adds an {\bf narrative space editor} on top of it (Fig.~\ref{system_interface}). The narrative space editor (Fig.~\ref{system_interface} (G, F)) supports editing the specification of the interactive story, including world description, character descriptions, and the rules. The narrative space visualizer takes the form of an infinite canvas of BSVs. Users can create, remove and edit BSVs, and reposition them on the canvas. We implemented all the BSV affordances described in Section \ref{sec:interactions}
%on top of the canvas
. Combining dimensions is achieved by dragging one BSV on top of another (Fig. ~\ref{system_interface} (C)). 
%Due to the visual complexity of higher-dimensional spaces, we restrict each BSV to have at most two dimensions. 
%We also only implemented categorical values for dimensions. To assist easy tracking of the temporal progression of any dimension, a predefined dimension called ``timeline'' is built in to be combined with any 1-dimensional BSV. 
The two narrative state highlighting mechanism described in Sec.~\ref{sec:highlighting_mechanism} for cross-referencing beteen multiple BSVs are triggered by the user clicking on individual narrative state %(Fig.~\ref{filtering_and_highlighting} (a))
and on a categorical value (combination). %(Fig.~\ref{filtering_and_highlighting} (b)).
The timeline control slider is located at the top of the canvas for global timestep filtering control (Fig.~\ref{system_interface} (L)).

Sometimes users need only a high-level overview of how narrative states distribute across different values of a narrative dimension, without temporal information---useful for assessing whether the value schema's granularity is appropriate by examining distribution evenness and per-value counts. We therefore introduce a compact version of the 1D BSV view (Fig.~\ref{system_interface} (K)),  that classifies narrative states by value without a timeline. Users can toggle between compact and timeline views via a button on each 1D BSV panel. The compact view also serves as an value schema editor (Fig.~\ref{system_interface} (K, bottom)).
%consists of three components: 1) {\bf Narrative space editor}, 2) {\bf Story playtester}, and 3) {\bf Narrative space visualizer} implementing our BSV visualization (Fig.~\ref{}). The narrative space editor (Fig.~\ref{}) supports editing the specification of the interactive story, including world description, character descriptions, as well as the rules. The story playtesting panel (Fig.~\ref{}) allows the user to play the interactive story based on the current specification. 

\system{} supports iterative authoring by allowing users to upload new batches of playthrough data (DG1) after modifying the story specification (Fig.~\ref{system_interface} (I)), retaining existing BSV narrative dimensions while updating the distribution of narrative states to reflect the new data. To facilitate before-and-after comparison, each BSV includes a toggle to show or hide the prior batch's distribution (DG3) (Fig.~\ref{trajectory_comparison}, Appendix). When real playthrough data is unavailable during early-stage development, users can instead prompt \system{} to generate simulated playthroughs (DG1) (Fig.~\ref{system_interface} (H)) via an LLM-based player proxy that models archetypes drawn from prior digital games research \cite{yannakakis2013player,worth2015dimensions} (Table~\ref{tab:player_archetypes}, Appendix). We present an example workflow in Sec.\ref{sec:example_workflow} in the Appendix.
%Table~\ref{tab:player_archetypes} (appendix) details player profiles implemented in \system{}.

\system{} uses Claude 3.5-Sonnet for both dimension extraction (concept induction) and dimension value classification, with a ReactFlow front end and Python Flask back end. We follow the concept induction pipeline of ~\cite{lam2024concept}, and implement the clustering operator by directly prompting the model to assign summarized narrative states into groups, for simplicity and reduced latency.
 %We leave the list of prompts we used in supplemental material.

\subsection{Example Workflow}
\label{sec:example_workflow}

This section presents an example workflow demonstrating the features described above. Alice is an interactive narrative writer who designs educational games for children around 10 years old. Alice wants to create an interactive story based on a fairy tale setting, aiming to teach kids that ``small act of courage can change everything.'' She opens \system{} and first types in a world setting and defines a few characters (Fig \ref{system_interface} (G)). In the story, the player plays as the duckling, who lives in a pond dominated by a bullying goose hoarding all the resources. The third character is the duckling's mother, who has a protective nature. Alice intends to convey the message by ensuring that player's courageous behavior leads to a happy ending, so she adds a rule saying ``\texttt{if the duckling stands up to the goose, all the ducks will join and fight together and reclaim the resources that belong to them.}'' Alice thinks this well communicates the message, so she prompts the system to generate the first batch of simulated playthroughs.

Alice notices that her newly added rule is rarely triggered during playthroughs, suggesting players seldom stand up to the goose. She suspects the duck Mom's generated dialogue discourages this behavior based on the character description. To test this hypothesis, she creates a narrative dimension called ``\texttt{whether\_the\_duck\_mom\_} \texttt{discourages\_the\_duckling}''.
%_from\_standing\_up\_to\_the\_goose}''.
The resulting BSV indicates this is the case for almost every narrative state. Alice thus revises the character description of the duck Mon to make her more encouraging. 

Alice also prompts the system to extract three key dimensions that differentiate the playthroughs from each other. The system returns ``\texttt{duckling\_behavior}'', ``\texttt{goose\_aggression\_behavior}'' and ``\texttt{location}''. Although lacking significant plot progression, Alice notices one common pattern: the goose's aggression behavior usually evolves from threatening to attacking as the story unfolds. This inspires Alice to add another rule to nudge the player towards more courageous behavior: ``\texttt{if the goose starts attacking, the duck Mom will get injured, and the duckling will become angry, thinking about confronting the goose.}''

Satisfied with the story, Alice generates a new batch of simulated playthroughs and notices a significantly improved rule triggering rate. To investigate how effectively the story communicates its message, she creates two BSVs ``\texttt{duckling\_courage}'' and ``\texttt{duckling\_happiness}'', and then combines them into a 2D view to examine their correlation.  The combined view confirms the intended pattern—courageous narrative states almost always lead to happiness by the end—but also reveals that the duckling can become happy without ever showing courage. Examining the data-derived dimension ``\texttt{location}'', Alice realizes that the duckling can escape and find another resource-rich pond, bypassing the intended arc. To eliminate such undesired storylines, Alice adds a new rule to restrict the story to the pond location ``\texttt{if the duckling attempts to leave the pond, they will realize that they have to stay at the pond.}'' Alice initiates another playthrough simulation. Comparing narrative state distributions across all iterations, Alice can clearly see that this modification brings her closer to effectively communicating her intended message (Fig.~\ref{bundled_branching_storyline}, bottom).

\section{User Study}
\revision{As an initial author-facing investigation, }we conducted a user study of \system{} to: 1) validate its support for perception and sensemaking of the narrative possibility space, 2) understand how perception of the narrative possibility space relates to control over space, and 3) explore how \system{} helps with creative exploration. 12 participants (4 female, 7 male, 1 non-binary; ages 23–37, mean 28) were recruited via relevant online communities. Participants had backgrounds in game design, AI/computer science/software engineering, and/or art. All with experience playing INs, 10 with experience playing AI-based  INs, and 7 had used AI in creating IN experiences. All reported at least moderate generative AI proficiency. \revision{6 of 12 are independent game developers with publicly released interactive narratives.} Each participant received a $\$100$ gift card. 
%in AI-based IN authoring.

%\subsection{Participants}

%Full demographic details are provided in the supplemental material.

%[[can expand]]

%\subsection{Conditions}

We compare \system{} with a baseline system where the BSV canvas is replaced by a playtesting panel that the author can employ during authoring---mirroring the setting of most existing LLM-based IN authoring tools~\cite{ai_dungeon2, inworldorigin, dramallama, wu2025orchid} (Fig~\ref{baseline} in Appendix). Each participant completed the authoring task twice, once with each system, with condition order counterbalanced across participants to mitigate order effects. Due to limited time and resources, participants worked only with simulated---rather than human---playthroughs in the \system{} condition. Each \system{} session therefore began by generating 3 simulated playthroughs from the participant's initial storyworld specification; participants then iteratively refined the specification and generated more simulated playthroughs as desired. To keep sessions timely, we capped {\em each} simulated playthrough at 5 rounds\footnote{Since participants can initiate simulation multiple times, the total rounds generated is always a multiple of 5.} and limited data-derived dimensions to 3. 

%\subsection{Task}

Participants were asked to select a theme from a provided list of 10 (Table~\ref{tab:story_themes}) and  create an interactive story (as described in Sec.~\ref{sec:interactive_stories}) using the provided authoring systems. They were instructed to limit their story to one scene and no more than four characters.
%To ensure the existence of a certain level of narrative intent, participants selected a theme for the story from a list of 10 themes we provided (Table~\ref{tab:story_themes} in Appendix) and were instructed to limit their story to one scene and no more than four characters.

\subsection{Procedure}

%A study session includes the following activities. 

\subsubsection{Set-up and Pre-study Survey (5 min and offline time)} 10 studies were conducted remotely via Google Meet, and 2 in person. Before each session, participants signed a consent form and selected a story theme for {\bf each} of the two conditions. At the start of the session, participants completed a pre-survey covering basic demographic information and their prior experience with Large Language Models, INs, and AI-based INs, followed by an introduction to the authoring task.

\subsubsection{Tutorial and Authoring Task (35 min each condition, 70 min total)} \label{sec:authoring_activity}
During each task, participants shared their screen and thought aloud. The facilitator introduced the interface and confirmed their understanding of key features through a brief pre-designed task. Participants then used the tool freely to create a story around their chosen theme, stopping when satisfied or when time ran out.

\subsubsection{Post-authoring Survey (5 min each condition, 10 min total)} \label{sec:post_survey_and_interview}
After each condition, participants completed a post-study survey (Fig.~\ref{post_study_survey}) covering self-evaluation of theme alignment, perception, control and player agency in their story (Fig.~\ref{post_study_survey} (a)) , along with questions adapted from NASA-TLX (Fig.~\ref{post_study_survey} (c)) and Creativity Support Index (Fig.~\ref{post_study_survey} (b)). For \system{} sessions, participants also rated the usefulness of individual components of the system (Fig.~\ref{post_study_survey} (d)). 

\subsubsection{Exit Interview (10 min)}
The study session concludes with a semi-structured interview in which participants provide open-ended qualitative feedback on both authoring systems and their AI-based authoring workflow more broadly.

\vspace{1mm}

%The activities described in Section \ref{sec:authoring_activity} and \ref{sec:post_survey_and_interview} were performed {\bf once} for {\bf each} of the two authoring conditions.
One complete study session takes roughly $95$ minutes plus offline time used for pre-study survey (Tab.~\ref{tab:study_procedure} in Appendix). 

\subsection{Results}

\begin{figure*}[t]
\centering
\includegraphics[width=0.90\textwidth]{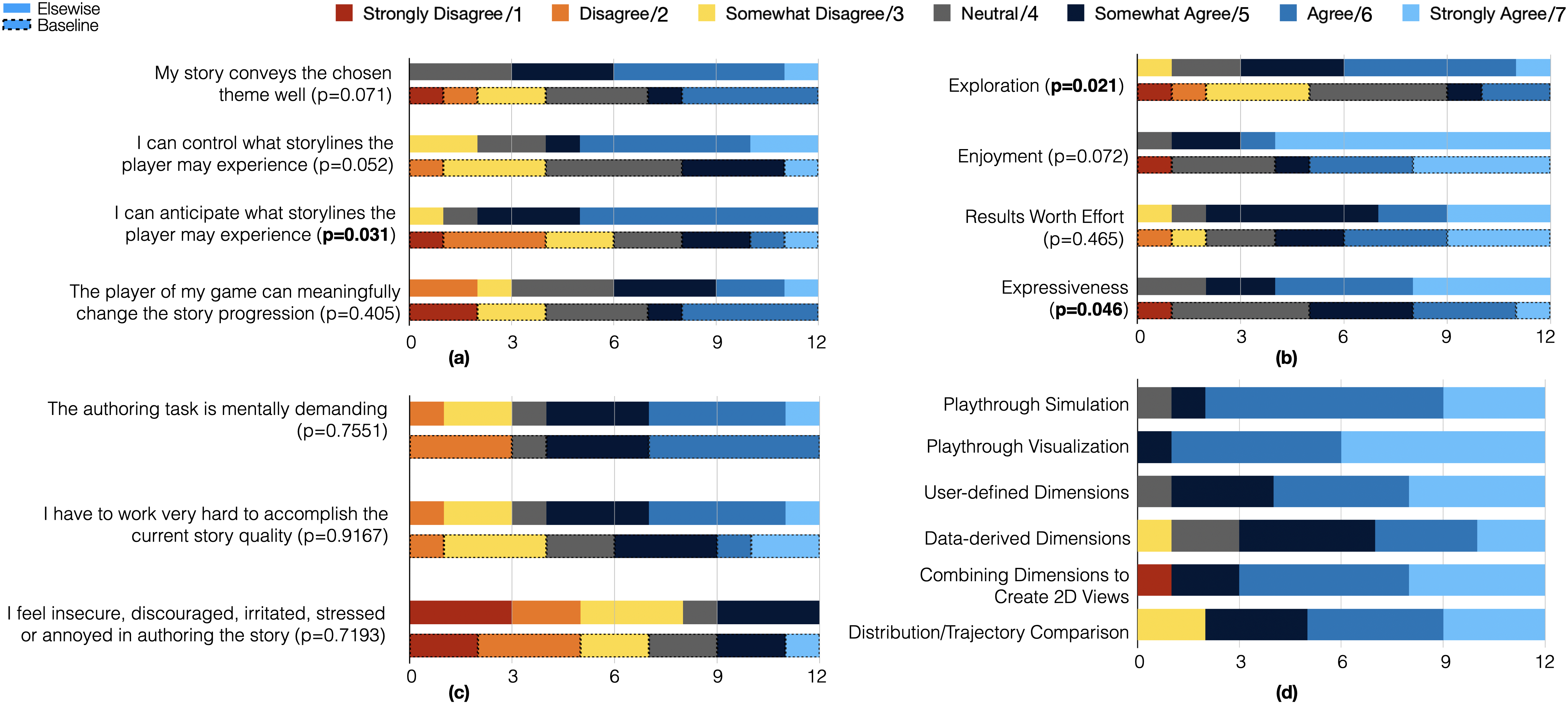}
\vspace{-10pt}
\caption{Post-study questionnaire results. (a) evaluation on theme alignment, control, anticipation and player agency comparing \system{} with the baseline; (b) Creativity Support Index related questions; (c) NASA-TLX related questions; d) Evaluation of usefulness of each \system{} component (1-extremely useless, 7-extremely useful)}
~\label{post_study_survey}
\Description{The figure presents comparative questionnaire results from a user study evaluating the Elsewise system against a baseline authoring tool. The data is displayed as stacked horizontal bar charts showing participant responses on a 7-point Likert scale from "Strongly Disagree" (1) to "Strongly Agree" (7), with borderless bars representing Elsewise and bars with dotted borders showing the baseline system. Panel (a) examines theme alignment, control, anticipation, and player agency. For "My story conveys the chosen theme well," Elsewise shows stronger agreement with more responses in the 5-7 range compared to the baseline. Similar patterns appear for controlling storylines and anticipating player experiences, where Elsewise users report better capability. For "The player of my game can meaningfully change the story progression," both systems show comparable positive responses, though Elsewise has slightly more strong agreement. Panel (b) displays Creativity Support Index metrics. "Exploration" and "Enjoyment" show marked better scores with Elsewise. "Results Worth Effort" and "Expressiveness" also favor Elsewise, with more responses in the agreement range and fewer in disagreement categories. Panel (c) presents NASA-TLX workload assessment questions. For "The authoring task is mentally demanding," "I have to work very hard to accomplish the current story quality," and  both systems show similar distributions centered around neutral to somewhat agree. For "I feel insecure, discouraged, irritated, stressed or annoyed in authoring the story," Elsewise demonstrates slight improvement with more responses in the disagree range, while both kind of had many disagreeing responses. Panel (d) evaluates specific Elsewise components' usefulness on a 1-7 scale from extremely useless to extremely useful. All features receive predominantly positive ratings above 4. "Playthrough Simulation," "Playthrough Visualization," "User-defined Dimensions" score highest with most responses at 5-7.  "Data-derived Dimensions," "Combining Dimensions to Create 2D Views," and "Distribution/Trajectory Comparison" showed some negative responses, but still the vast majority responded positively.}
\vspace{-15pt}
\end{figure*}

We collected survey responses, semi-structured interview recordings, and interaction logs from the interface. To examine differences between the two authoring conditions, we applied the Wilcoxon signed-rank test to post-study survey results. Interview audio was transcribed and analyzed using thematic clustering~\cite{blandford2016qualitative}. \revision{Thematic analysis was conducted by one author and validated by co-authors.}

Post-study survey results are shown in Fig.~\ref{post_study_survey}. Participants rated \system{} significantly more positively than baseline across anticipation, exploration, and expressiveness, and reported enjoying the authoring process overall. Mental effort was rated slightly higher for \system{}, though the difference was not significant. Negative feedback centered on a steep learning curve, particularly around visualization complexity of 2D BSVs. On average, participants created 4 rules, 6.33 BSVs, and 1.75 2D BSVs with \system{}, compared to 3.42 rules with baseline. \system{} sessions involved an average of 3.6 authoring iterations (each defined as modifying story specifications followed by playthrough simulation), and participants explored an average of 54.4 playthrough rounds with \system{} versus 17.2 with baseline. Complete BSV canvases for each participant are provided in the supplemental material. %(Tab. \ref{tab:authoring_process_statistics}).

% \begin{table}[b]
% \small
% % \resizebox{\linewidth}{!}{
% \caption{Statistics on Authoring Process}
% \begin{tabular}{| p{5cm} | p{3cm} | p{3cm} | }
% \hline
% {\bf Metrics} & {\bf \system{}} & {\bf baseline} \\ \hline
% Avg. Rule Created & 4 & 3.42 \\ \hline
% Avg. Playthrough Rounds Explored & 54.4 & 17.2 \\ \hline
% Avg. BSV views Created & 6.33 & - \\ \hline
% Avg. 2D BSV views Created & 1.75 & -  \\ \hline
% \end{tabular}
% \vspace{2pt}
% \Description{}
% \label{tab:authoring_process_statistics}
% % \vspace{-25pt}
% \end{table}

%[[behavior data]]

\subsubsection{\system{} assists authors to better perceive the narrative possibility space} Participants reported greater confidence anticipating 
%potential
storylines with \system{} ($Med: 6$) than baseline ($Med: 3.5$, $p = 0.031$). We identify two aspects through which \system{} helps authors better perceive the narrative possibility space.
\vspace{-1mm}
\paragraph{Providing a player perspective} 10 of 12 participants explicitly appreciated \system{}'s incorporation of player perspectives into the authoring process, noting that it helps to overcome the {\bf biased author perspective}
%when playtesting}
(P4: ``\textit{When I'm playtesting I'm intentionally triggering the key plot points, which may not be what an actual player would do}'' and P5: ``\textit{As the author, I can't erase my memory about the rules I added.}''). The lack of player perspective matters less in traditional IN authoring, where possible actions are predefined and finite. In open-ended IN, {\bf the seemingly infinite player action space makes authoring feel overwhelming}  (P7: ``\textit{There seems to be infinite possibilities. I'm kinda not sure how to proceed to write rules. I want to at least know what the characters can do.}'') and make the resulting story hard to debug (P4: ``\textit{This types of games %(AI-based INs) 
gives too much agency to players. Even traditional INs 
where players can only select from a limited set of options 
is hard to debug.}''). Although participants know the simulated player behaviors are inaccurate, they still appreciate that different player profiles {\bf diversify the experienced narrative through varied behavioral patterns} (P5: ``\textit{It gives me the means to see how a story prompt may play out across a number of different player types.}''), which helps with ``\textit{finding edge-cases in story writing}''(P7) and exploring different possible storylines (P12: ``\textit{it's like exploring different universe}''). P4 directly noted ``\textit{AI players may be inaccurate in representing real players, but they help in the sense that they enumerate all the possible player behaviors}''. P8 suggested allowing user-defined player profiles. 
%Notably, we observed that some participants overfit to a single simulated profile, crafting rules solely to counter the killer player (P5).
\paragraph{Better sensemaking of complex storylines with BSV} Participants expressed challenges in making sense of playthrough data during authoring with our baseline tool (P5: ``\textit{I don’t have a sense of where things change and where is plot progression}'') Playthrough visualization received highest median and average {\em usefulness} rating among all components (Fig.~\ref{post_study_survey} (d)). 6 out of 12 participants explicitly expressed their strong appreciation of the visualization.
%(e.g., P8:``\textit{I'd give it 9 point (Likert score) if I had the option}''). 
Specifically, participants find \system{} useful in the following aspects:
%\vspace{-1mm}
\begin{itemize}
    \item {\bf Outlining plot progression} in varied aspects (P5: ``\textit{it gives you a thematic overview, especially..how the story progresses and what the players may potentially feel from all of it}'' ),
    \item {\bf Systematically summarizing narrative possibilities} across dimensions (P4: ``\textit{It feels like a heatmap for plot possibilities. It’s quite interesting.}'', P7: ``\textit{ [combining dimensions] helps to investigate the narrative in a nuanced way}'')
    \item {\bf Connecting narrative intents to concrete plot.} 
    %by creating semantic dimensions. 
    P6 noted ``\textit{this tool helps me to convert the abstract theme I want to express to something more concrete}'' and P7 expressed that ``\textit{defining the semantic and see how different playthroughs reside in them allows me to see how alternative versions of the story adapt to same concepts differently}.'' (P7).
\end{itemize}
%\vspace{-1mm}
Most criticism centered on the visual complexity of 2D BSVs---P6 and P8 explicitly struggled to interpret them. P4 raised concern on potential mismatch between the author and the LLM in how dimensional meaning is understood.

%[[Variable created]]
%[[Automatically extracted variables utilized]]

%``It feels like it's two different tasks - for the baseline I was trying to debug my rules, for Elsewise I was taming the player.''

\subsubsection{How perception of the narrative possibility space helps with controlling the space} \label{sec:control_result}

Although \system{} does not offer a distinct control scheme for shaping the narrative possibility space, participants nonetheless reported feeling greater authorial control when using it ($Med: 6$) than baseline ($Med: 4$, $p=0.052$). We identify the following possible reasons.
%participants believed \system{} offers better control.
\vspace{-2mm}
\paragraph{Better perception gives more confidence in control} Many participants struggled to control the system due to difficulty perceiving the effects of prompt changes. (P2, P3, P4, P5, P6, P7, P8). In our context, as P5 noted, ``\textit{it’s hard to distinguish what rules will guide the player towards
a desired outcome, though I think that’s the case for
all AI authoring in general}''. Participants appreciate the ``\textit{tightly coupled data visualization, simulation and rule/story authoring}'' (P4), which enables rapid trial-and-error loops of ``\textit{defining where you are guiding the player to and seeing where it actually goes.}'' (P5) P7 reported that comparing distribution across two iterations ``\textit{helps me to examine naunced difference across different versions led by small prompt changes}''. Participants mentioned that \system{} makes them``\textit{more motivated with quick feedback/positive signal}'' (P6, P10). In the end, P6 explicitly noted, ``\textit{I’m more confident in my control after seeing so many playthroughs. Even though I know it won’t always unfold like how I expect, I think it's ok. A lot of unexpected situation are just outliers and they don’t affect my confidence.}'' In contrast, participants using the baseline---despite having the same control mechanism---sometimes stopped the authoring process early, lacking a clear sense of how their efforts were affecting the outcome (P6 and P11).
%(P6: ``\textit{I don't know how to meaningfully proceed}'', and P11: ``\textit{I'm not sure when it can be considered as `done'}''). 

\begin{figure}[t]
\centering
\includegraphics[width=0.45\textwidth]{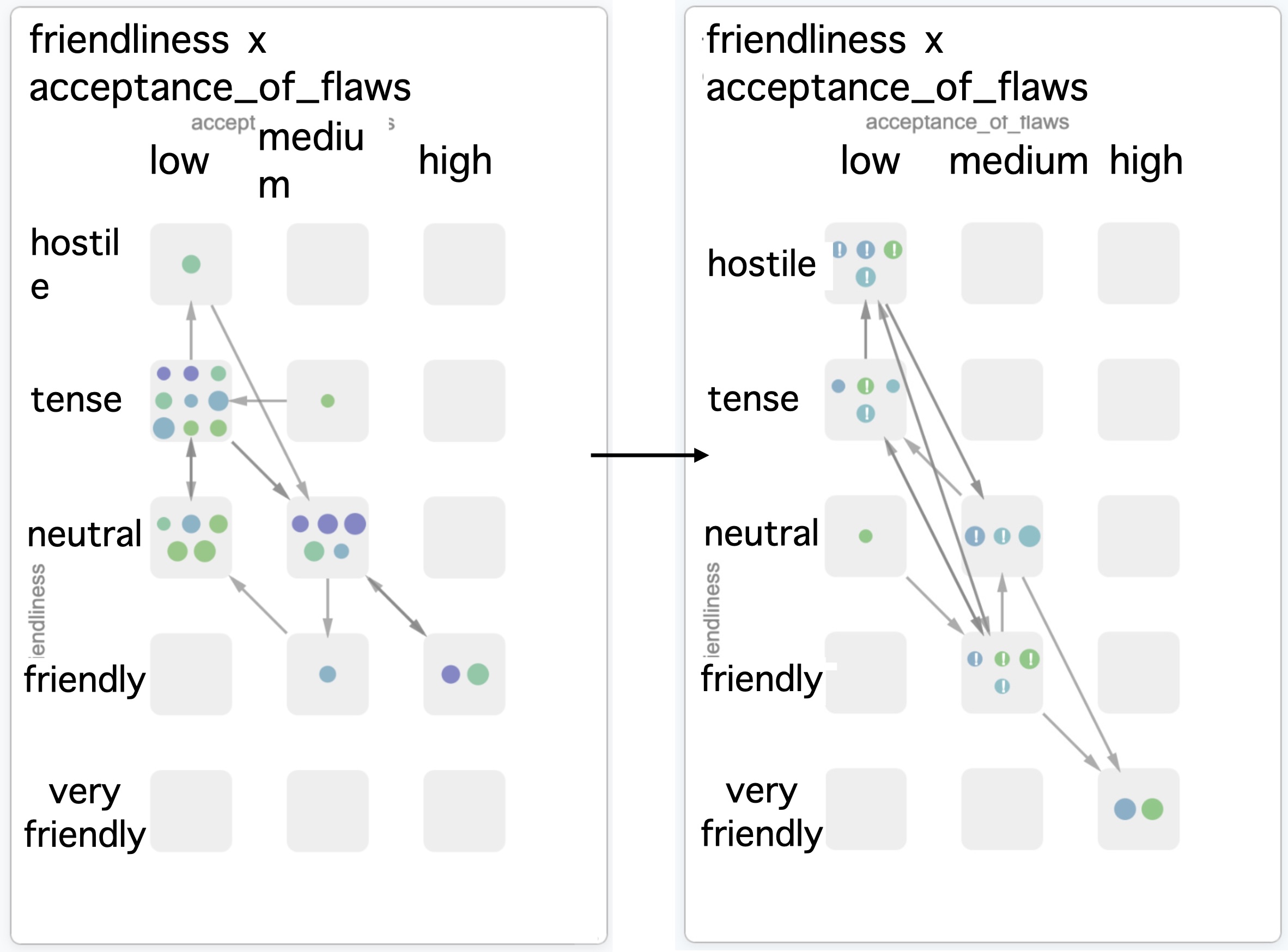}
\vspace{-10pt}
\caption{P4 attempted to strengthen the positive correlation between friendliness and acceptance of flaws to communicate ``\textit{True friendship means accepting someone's flaws}''. 
%; (c) P10, ``\textit{The journey matters more than the destination}''
}
\Description{This figure demonstrates how the Bundled Storyline Visualization helps authors understand and refine narrative state distributions through rule authoring, showing before and after comparisons for two participants working with different thematic goals. The example shows Participant 4's attempt to strengthen the correlation between friendliness and acceptance of flaws to convey the theme "True friendship means accepting someone's flaws." The left visualization displays the initial state distribution across a 2D grid combining friendliness levels (hostile, tense, neutral, friendly, very friendly) with acceptance of flaws (low, medium, high). Colored dots represent narrative states, with connections showing possible story trajectories. The initial distribution shows scattered states without a clear positive correlation between the two dimensions. After rule authoring, shown on the right with an arrow indicating the transformation, the distribution shifts notably. More narrative states now cluster along a diagonal pattern from hostile/low acceptance to very friendly/high acceptance, demonstrating a stronger positive correlation. The connecting lines between states also show more coherent progression paths that reinforce the intended theme.}
~\label{trajectory_improvement_examples}
\vspace{-25pt}
\end{figure}

\vspace{-2mm}
\paragraph{Externalization of intermediate control objects} The BSVs give authors tangible objects to manipulate, allowing them to intuitively shape the distribution of narrative states rather than reasoning abstractly about possible storylines
%—the ``\textit{potential perspectives}'' (P6)---
(Fig.~\ref{trajectory_improvement_examples}, and Figure~\ref{trajectory_improvement_examples_all} in Appendix). For example, when looking at the BSV of the dimension ``\texttt{genre}'', P4 started to imagine ``\textit{first comedic, then dramatic, and then conflict---kinda like designing a desired process}''. P8
%, who directs interactive theater, 
noted that 
%such trajectories (e.g., emotional arc, character development) are typically implicit, and 
visualizing such trajectories (e.g., emotional arc, character development) helps ``\textit{communicate them well externally}''. The ability to visualize narrative states along arbitrary semantic dimensions led some participants to manipulate highly intangible story aspects---P10 created a dimension called ``\textit{likelihood\_of\_buying\_the\_game}'' and attempted to guide most storylines towards the ``\textit{very\_likely}'' value at later stages on timeline. (Fig. ~\ref{trajectory_improvement_examples_all} (c), Appendix) 
% Could move Figure~\ref{trajectory_improvement_examples} to appendix
\vspace{-2mm}
\paragraph{Data-grounded level of abstraction in prompting language} When participants viewed the distribution of narrative states as the object of control, some began using more abstract language in their rules to achieve broader impact on the narrative space. For example, P1 first wrote a rule ``\textit{if the cat helps the mouse, then they become friends}'', then realized that what they really meant was ``\textit{if two characters help each other, then they become friends.}'' 
%P4 also explicitly asked ``\textit{can I be vague in writing the rules?}'' 
%Extreme examples include P5 writing the rule ``\texttt{if the story strays from the theme, it guides the characters back to more relevant plotlines}''. 
When narrative dimension values are derived inductively from data, they can guide participants in crafting prompts that precisely enumerate possibilities,
%within each dimension
as observed when participants drew directly on these generated values when writing rule conditions (P1, P4).

\subsubsection{How does \system{} help with creative exploration in AI-based IN authoring}\label{sec:exploration_result}

Although not designed as an ideation tool, participants reported that it's easier to explore ideas with \system{} ($Med: 5.5$) than baseline ($Med: 4$, $p=0.021$), and they feel more expressive and creative using \system{} ($Med: 6$) than baseline ($Med: 5$, $p=0.046$). We provide relevant qualitative findings below.
% This can be explained from the following aspects.
\vspace{-2mm}
\paragraph{More efficient exploration of (experienced) storylines} Concrete examples of simulated storylines inspired authors in designing key plot points, steering narrative direction, and nudging player behavior. For instance, P5 deliberately generated many playthroughs to brainstorm ways to convey the theme. P6, despite having initial ideas for key plot points, ``\textit{forgot what I had in mind when I look at the simulated playthrough}'' and pivoted to a different strategy. P8 was struck by some ``\textit{unexpectedly interesting plot}'' that emerged from one of the simulations. Note that the baseline also inspired participants through concrete storylines, though far less frequently—due to less efficient exploration. On average, participants experienced $54.4$ rounds of simulated playthrough with \system{}, compared to only $17.2$ rounds with the baseline. Accordingly, 9 out of 12 participants explicitly preferred \system{} for exploring possible storylines. One interesting exception is P10, who prefers baseline in terms of exploration because ``\textit{it (baseline) promotes more active and fine-grained thinking because I’m experiencing the story myself.}'' and ``\textit{I have more control in debugging my rules (because I can actively trigger them as the player)}''.

\vspace{-2mm}

\paragraph{More efficient exploration of relevant dimensions} The data-derived dimensions sometimes surface aspects the author had overlooked. For example, P4 noted ``\textit{The extracted dimensions feel quite interesting...maybe I should have different branches of the plot line to be in different genre?}''. Displaying all value combinations in each BSV systematically surfaces unexplored possibilities, naturally prompting reflection on uncovered cases (P5: ``\textit{It’s interesting that someone could be courageous but having low acceptance of the situation}''). P4 also remarked, ``\textit{although there is no playthrough data in the bucket of hostile x high (acceptance), it makes me think what would happen if this is the case.. There could be a branching storyline there. It’s very clear what are all the possibilities}''.

\vspace{-2mm}

\paragraph{Exploration is fun and engaging with \system{}} Participants were notably more motivated to explore the story space when using \system{}, likely due to its quick, intuitive feedback and ability to probe arbitrary narrative aspects. P6 said ``\textit{(this tool) makes me curious and excited about the progression of the story}''. P8 also noted ``\textit{the data visualization makes the exploration fun}''. 
%When asked for other potential use cases of the tool, P10 put ``\textit{for fun}'' as one of the use case. P4 requested to use \system{} to analyze their baseline-created story after the formal session purely out of curiosity. 
Consistent with these observations, enjoyment scores on our post-study questionnaire (Fig. \ref{post_study_survey}) reached a median of 7 out of 7 on the Likert scale.

% \subsubsection{How does IN authors incorporate BSV interactions in their authoring workflow}
% %[[additive vs. subtractive design process]]

% [[BSV does not introduce different control mechansm, however it changes how user wants to control the narrative space (designing the process)]]

% [[the sense of control originates from both actually restricting the player's action through rules, and having a clear sense of what aspect of the story actually matters ]]

% [[learning curve]]

% [[Survey question (control)]]

% [[Key plot points first; then nudging]]

% [[theme-related vs. non-theme related rules/dimensions]]

% [[How dimensions inspire rule creation]]

% [[Compare abstraction level and types of rules from DM vs EW (key plot points, nudging vs. steering)]]

% [[Quotes]]

% [[Limitation]]

% [[Survey question]]

% [[Measurement on plot progression]]

% [[Trajectories under the same theme as qualitative evidence]]

% [[Quotes]]
\section{Discussion}
We discuss implications from our findings and future directions.
%Authoring AI-based INs presents unique challenges compared to conventional INs—particularly around perception, control, and exploration within the narrative possibility space. While the Bundled Storyline concept and \system{} offer a preliminary approach to these challenges, new tools and paradigms remain worth pursuing.
%We discuss implications from our findings and future directions.

\subsection{Shaping an unknown open-ended possibility space}

Interactive narrative authoring is inherently challenging: the final experience remains incomplete until audience interaction occurs, yet the audience is absent during authoring. Authors must therefore work under assumptions about potential audience behaviors. 
%While AI-based interactive narratives offer exciting possibilities for new media forms, the blackbox nature of current AI methods introduces additional uncertainty into the co-creation formula. 
AI-based open-ended INs compound this challenge---the blackbox nature of current AI methods introduces additional uncertainty into the co-creation process. 
Both \system{} and baseline unsurprisingly received high mental effort ratings in our study (Fig.~\ref{post_study_survey} (c)). As P4 mentioned (``\textit{the mindset required for authoring this kind of interactive narratives is very different (from traditional IN).}''), this difficulty is partly fundamental to the authoring task itself.

Conventional narrative design works {\em additively}---authors create content to expand a possibility space.  In contrast, LLM-based interactive narrative design requires authors to {\em subtractively} eliminate undesired content from an initially boundless possibility space. (P4: ``\textit{I start to think about restricting the possibilities}''). This requires the author to first have some idea of ``what is already there'' to even start authoring. P7 explained their thought process when authoring with the baseline ``\textit{because I don’t know the specific possibilities, I need to be very comprehensive, I basically first created an action schema in my head for the characters and enumerate one by one all the possible combinations}''. This self-imposed ``action schema'' serves a dual purpose—tracking authorial progress while demarcating the boundary between authored territory and the "wilds" governed entirely by the LLM and player interactions.

LLM-based IN authoring tools could tackle these challenges by:
\vspace{-2mm}
\paragraph{Presenting the narrative possibility space in a perceivable and navigable way.}
Besides Bundled storylines, other approaches could help authors understand ``what's already there'' before subtracting or constraining possibilities. For example, an authoring tool could answer targeted queries by simulating story outcomes (e.g., ``show me 10 ways this conversation escalates into conflict'').
\vspace{-2mm}
\paragraph{Assisting the author in effectively shaping the narrative possibility space.}
Our system gives authors no direct control over the narrative possibility space. When storylines drift into undesired territory, an authoring tool could generatively suggest revisions to the author prompts---nudging future storylines toward intended regions.
\vspace{-2mm}
\paragraph{Supporting progressive authoring through structure and scaffolding} 
Rather than fully open-ended prompts, an authoring tool could offer structured templates to help authors externalize their mental model of the story, visually distinguishing between "authored" and "AI-generated" territory to clarify boundaries and progress. To reduce cognitive load, the tool could scaffold authoring into discrete phases—establishing the storyworld, exploring possibilities, designing key events, and incrementally building and debugging constraints—while automatically detecting underconstrained spaces and surfacing targeted questions. This lets authors grow narrative complexity gradually
%, validating constraints at each stage 
rather than specifying everything upfront.

\subsection{Player agency in LLM-based IN}

\revision{A primary advantage of LLM-based interactive narratives is their capacity to provide virtually unlimited player agency. Interestingly, when we ask our participants to evaluate the player agency offered by their created interactive stories (Fig.~\ref{post_study_survey} (a)), their assessments were more moderate than theoretical expectations might suggest (Med: 4, across both authoring conditions). Despite the system's free-form input capabilities and dynamic response generation, participants remained cautious in attributing high levels of genuine player agency to their narratives (P4: ``\textit{the freedom a chatbot gives to a player doesn’t necessarily equal to player’s freedom in the game...player can play endlessly, but there is no plot progression.}'') }

\revision{When asked about reasons behind their ratings, P7, P8, and P10 also acknowledged that their ratings are relative to the intended storylines they have authored. P10 articulated that when considering this question, they mentally constructed a branching storyline structure based on their authored rules and then evaluated how much freedom this structure offered players.}

%P2 responded with a very low score on player agency (2), saying that ``\textit{my story is completely linear}''. Their story contains one key plot event that is almost guaranteed to happen regardless of player actions. They appeared to view the occurrence of this key plot event as the only meaningful direction for plot progression. 

\revision{These findings align with existing theories of player agency, particularly with the recognition that permitting actions within a system does not necessarily confer meaningful agency if those actions lack support from the underlying computational model~\cite{wardrip2009agency}. In conventional interactive narratives, authors usually possess comprehensive understanding of the computational model, enabling them to deliberately shape player agency within their designs. AI-based interactive narratives present a fundamentally different situation: the opaque nature of the underlying computational model obscures this comprehension not only from players, but also authors. Consequently, authors cannot fully determine what constitutes "meaningfully supported" actions, and their conceptual models of agency may diverge significantly from player experience—particularly because emergent interactions between players and the system now constitute a substantial source of perceived meaning. This divergence highlights potential disparities between how authors and players conceptualize agency in AI-based interactive narratives and suggests the need for authoring tools that foreground player perspectives on agency affordances to reconcile both parties' conceptual models of meaning within the game.  }

whether player agency can play any meaningful role in author's artistic expression.

%\revision{Admittedly, the lack of real player data is a limitation of our user study. A clear next step is to evaluate the system with actual players and obtain better understanding about player-perceived agency. }

\subsection{Implications for AI agent design support}

While this paper discusses bundled storylines within digital games, the framework in theory applies to any interactive medium using runtime response generation---including AI agents built for \revision{open-ended} audience interaction, \revision{human-in-the-loop task solving} and information exchange. Designers of such systems must perceive, control, and explore the possibility space of user experience (\cite{tang2025understanding}). In this work, the AI agent acts as a game master, translating the author's narrative vision into responses to player actions\revision{, a pattern echoed in systems where an agent narrates or reacts to a player's actions in service of an authored goal, such as in \cite{zhang2025can} and \cite{drago2025improvmate}}. \revision{More broadly, the approach could support general conversational agent design, as in prompting-based chatbot construction~\cite{zamfirescu2023herding}, community-driven, case-based agent design~\cite{kuo2026botender}, narrative-based learnersourcing~\cite{moore2024teaching}, or AI-based design assistants for game creation~\cite{earle2025dreamgarden}}.
%, or generative user interfaces in general
\revision{In some of these settings, rather than an author dictating responses, multiple parties could use the agent together to simulate and explore outcomes. This points to a collaborative sensemaking mode that we leave to future work.}

%More broadly, the approach could power virtual customer assistants, virtual therapists, or even generative user interfaces.

While integrating a bundled storylines-based possibility space visualizer into a general AI agent builder~\cite{openai_agent_builder, n8n} is conceptually feasible, several challenges remain: identifying meaningful semantic dimensions, designing effective visualizations, and adapting control affordances to specific domains. Task-driven applications pose additional difficulties—representing functional behavior as semantic dimensions, reconciling explicit system states with semantic representations, distinguishing hard constraint violations from soft constraint deviations, and simulating user behavior from real-world data. We hope our initial investigation in the context of digital games opens the door to possibility space visualization as a broader design support tool for general AI agents.

%Furthermore, the initial possibility space based on an LLM model isn't just a piece of marble passively waiting to be sculpted - it has its own narrative tendency that the author needs to actively tame or even fight against.

%[[Applications outside IN]]

\section{Limitation}
We summarize the major limitations of our work as follows:

\vspace{-2mm}

\paragraph{Player Experience} Due to time and resource constraints, this work focused primarily on the author's experience, with player involvement limited to LLM-simulated playthroughs. Feedback from real human players would yield more accurate insights into narrative intent communication and player-perceived agency. \revision{Lack of real player data in our study is a clear limitation and player-facing study is a clear next step.}

\vspace{-2mm}

\paragraph{Study Design} Story scale and playthrough data were restricted to sizes smaller than most real-world use cases to keep sessions within time limits; a longitudinal study with more realistic projects remains future work. Additionally, evaluating the system with more diverse groups---particularly those with limited AI or IN experience---would better illuminate its usability and generalizability.

\vspace{-2mm}

\paragraph{Interface Design} The interface design could be further improved, particularly to reduce the visual complexity of 2D BSVs. \revision{Visual simplification and distinctions can be applied to improve the clarity of 2D BSVs.} Usability would also benefit from additional features such as user-defined player profiles, customizable dimension extraction granularity, as well as support for non-categorical dimension values.

\vspace{-2mm}

\paragraph{Technical Method} \revision{Our definition of bundled storylines presupposes a global timeline, shared across storylines and divisible into discrete timesteps—raising the unaddressed question of how to group narratively equivalent event sequences unfolding at different paces. Techniques like dynamic time warping or temporal semantic alignment offer potential solutions, but we leave this to future work, since defining semantic similarity at the level of narrative dimensions, rather than surface text, remains an open problem in itself.} Additionally, as this work focuses on novel interface design, we implemented a simplified, unevaluated version of concept induction~\cite{lam2024concept} for dimension extraction, though alternative strategies may 
affect
%the granularity of extracted concepts and, in turn, 
authors' workflows and resulting story specifications.   %, though the success of LLMs at inferring and making use of similar semantic dimensions in other systems

\section{Conclusion}
We introduced Bundled Storylines, a novel concept for comprehensive sensemaking of AI-based interactive narratives, and implemented it in the authoring tool \system{}. Bundled Storylines enables IN authors to intuitively shape the narrative possibility space across user-configurable dimensions. Our \revision{initial author-facing investigation} demonstrated its effectiveness in helping authors anticipate player-experienced storylines and better control the narrative possibility space. Our work advances AI-based IN authoring workflows towards control and balance between authorial intent and player agency.

\bibliographystyle{ACM-Reference-Format}
\bibliography{bib}

\appendix
\clearpage
\section{Appendix}
\subsection{Additional illustrations of Bundled Storylines and Affordances }
%Before having any specific idea on how to convey the message, Alice first uses the playtesting panel to get a feel of possible way the story could unfold. She played the game 4 times. In the four playthroughs, the characters seem to be just showcasing their character description in different ways, without actually pushing the plot forward towards any direction that Alice could recognize. At each playthrough Alice soon starts to feel repetitive. To understand if there is any meaningful development in the four playthroughs, Alice prompts the system to extract three key dimensions that help differentiate each narrative states from the others from the 4 playthroughs. (Fig~\ref{}) From the extracted dimensions, Alice noticed one common patterns across the 4 playthroughs: the goose's aggression level usually evolve from threatening to attacking as the story unfolds.

\begin{figure*}[bth]
\centering
\includegraphics[width=\textwidth]{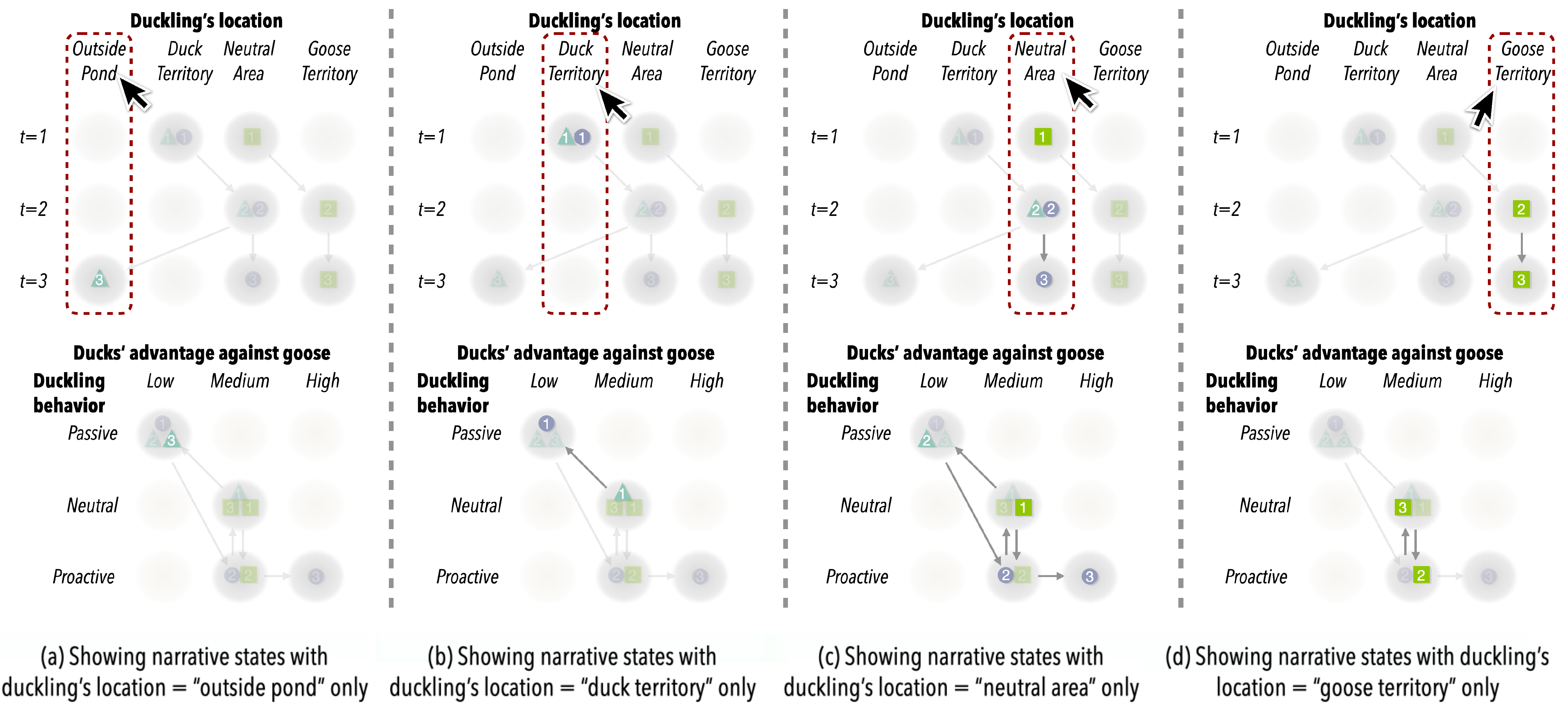}
\vspace{-10pt}
\caption{Multiple BSV views with dimension
value filter. Red dashed boxes indicate the selected dimension values, and the bottom BSV reflects how other BSV views on the canvas update in response to the filtering operation. The example reveals a correlation between duck's advantage over goose and duckling behavior across all possible duckling locations: more proactive duckling behavior increases the likelihood of duck's advantage when the duckling is in a neutral area, but not in goose territory.}
\Description{Four sub-figures (a) through (d), each showing the same two-part state-transition diagram filtered to a different duckling location: Outside Pond, Duck Territory, Neutral Area, and Goose Territory respectively. In each sub-figure, a red dashed box with a black arrow highlights the active location column; inactive states are faded. Colored dots within each gray bubble indicate which storylines are active in that narrative state, with blue, teal, and green each representing a distinct storyline. Each sub-figure has a top grid with rows for timesteps t=1, t=2, t=3 and columns for the four locations, and a bottom grid with rows for duckling behavior levels (Passive, Neutral, Proactive) and columns for ducks' advantage against goose (Low, Medium, High). Sub-figure (a), Outside Pond: the top grid shows a state with the teal storyline appearing at t=3 in the Outside Pond column, with diagonal transition lines connecting states in other locations at earlier timesteps. The bottom grid shows an active state with the teal storyline at Passive/Low, with a diagonal line descending to Neutral/Medium and continuing to Proactive/High. Sub-figure (b), Duck Territory: the top grid shows a state with the teal and blue storylines at t=1 in Duck Territory, with a diagonal transition line to Neutral Area at t=2. The bottom grid shows a state with the blue storyline at Passive/Low, with an arrow pointing diagonally down toward Proactive/Medium-High. Sub-figure (c), Neutral Area: the top grid shows a state with the green storyline at t=1, a state with the blue and teal storylines at t=2, and a state with the blue storyline at t=3 in Neutral Area, connected by a downward arrow from t=2 to t=3. The bottom grid shows a state with the teal storyline at Passive/Low, arrows leading to Neutral/Medium (green and teal storylines active), a bidirectional arrow to Proactive/Medium, and a rightward arrow to Proactive/High (blue storyline). Sub-figure (d), Goose Territory: the top grid shows a state with the green storyline at t=2 and t=3 in Goose Territory, connected by a downward arrow. The bottom grid shows a state with the green storyline at Neutral/Medium, a bidirectional vertical arrow to Proactive/Medium, and a diagonal line extending toward the Passive level.}
~\label{multiple_bsv_with_dimension_value_filter}
\end{figure*}

\begin{figure}[tbh]
\centering
\includegraphics[width=0.5\textwidth]{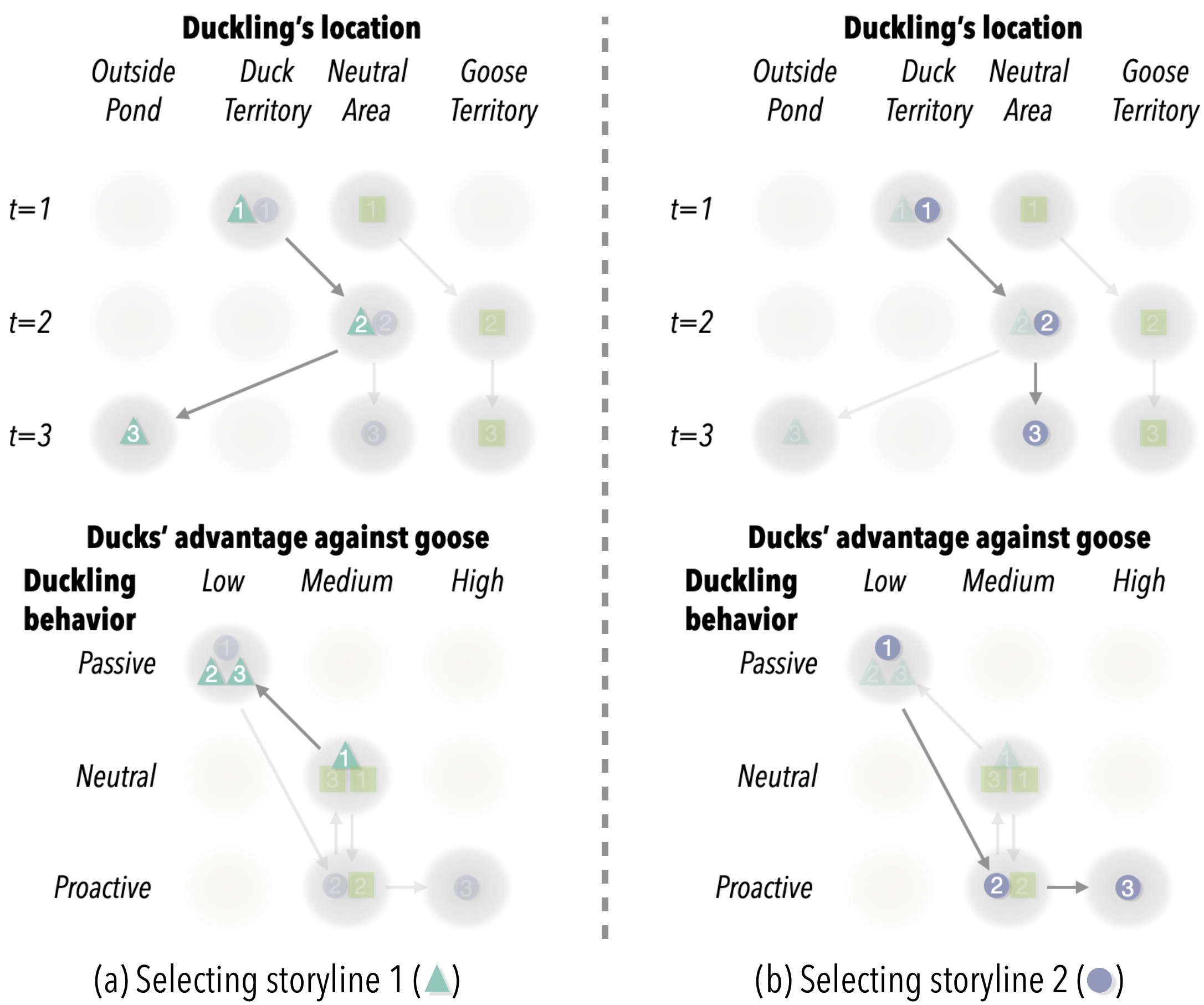}
\caption{Multiple BSV views with storyline filter. In the blue (first) storyline, the ducks are displaced outside the pond (top), consistent with their declining advantage over the goose and increasingly passive duckling behavior (bottom); the purple (second) storyline exhibits the opposite trend. }
\Description{Two side-by-side BSV diagrams, labeled (a) and (b), each filtered to highlight a single storyline. Colored dots within gray bubbles represent narrative states belonging to different storylines; inactive states from unselected storylines are faded. Each sub-figure has a top grid showing Duckling's Location (columns: Outside Pond, Duck Territory, Neutral Area, Goose Territory) across three timesteps (t=1, t=2, t=3), and a bottom grid showing Duckling behavior (rows: Passive, Neutral, Proactive) against Ducks' advantage against goose (columns: Low, Medium, High). Sub-figure (a) selects storyline 1, represented by a teal dot. In the top grid, a black arrow highlights the starting state at t=1 in Duck Territory, containing a teal dot. An arrow leads diagonally to t=2 in Neutral Area, where a teal dot appears alongside a faint yellow-green dot. A further arrow leads to t=3 in Outside Pond, where only the teal dot remains, indicating the duckling has moved outside the pond by the final timestep. In the bottom grid, the Passive/Low state contains a teal and yellow-green dot; an arrow leads diagonally down-right to a Neutral/Medium state with a teal and small green dot; a bidirectional arrow connects to a Proactive/Medium state; and a rightward arrow leads to a Proactive/High state. This reflects a trend of declining advantage against the goose and increasingly passive duckling behavior. Sub-figure (b) selects storyline 2, represented by a blue (purple) dot. In the top grid, a black arrow highlights the starting state at t=1 in Duck Territory, containing a blue dot. A downward arrow leads to t=2 in Duck Territory (blue dot), then another downward arrow to t=3 in Duck Territory (blue dot), showing the duckling remains in Duck Territory throughout. In the bottom grid, the Passive/Low state contains a blue and teal dot; an arrow leads diagonally down-right to a Proactive/Medium state containing a blue and green dot; a bidirectional arrow connects within that level; and a rightward arrow leads to a Proactive/High state with a blue dot. This reflects the opposite trend: increasing advantage and increasingly proactive duckling behavior.}
~\label{multiple_bsv_with_storyline_filter}
\vspace{-15pt}
\end{figure}

\begin{figure}[tbh]
\centering
\includegraphics[width=0.5\textwidth]{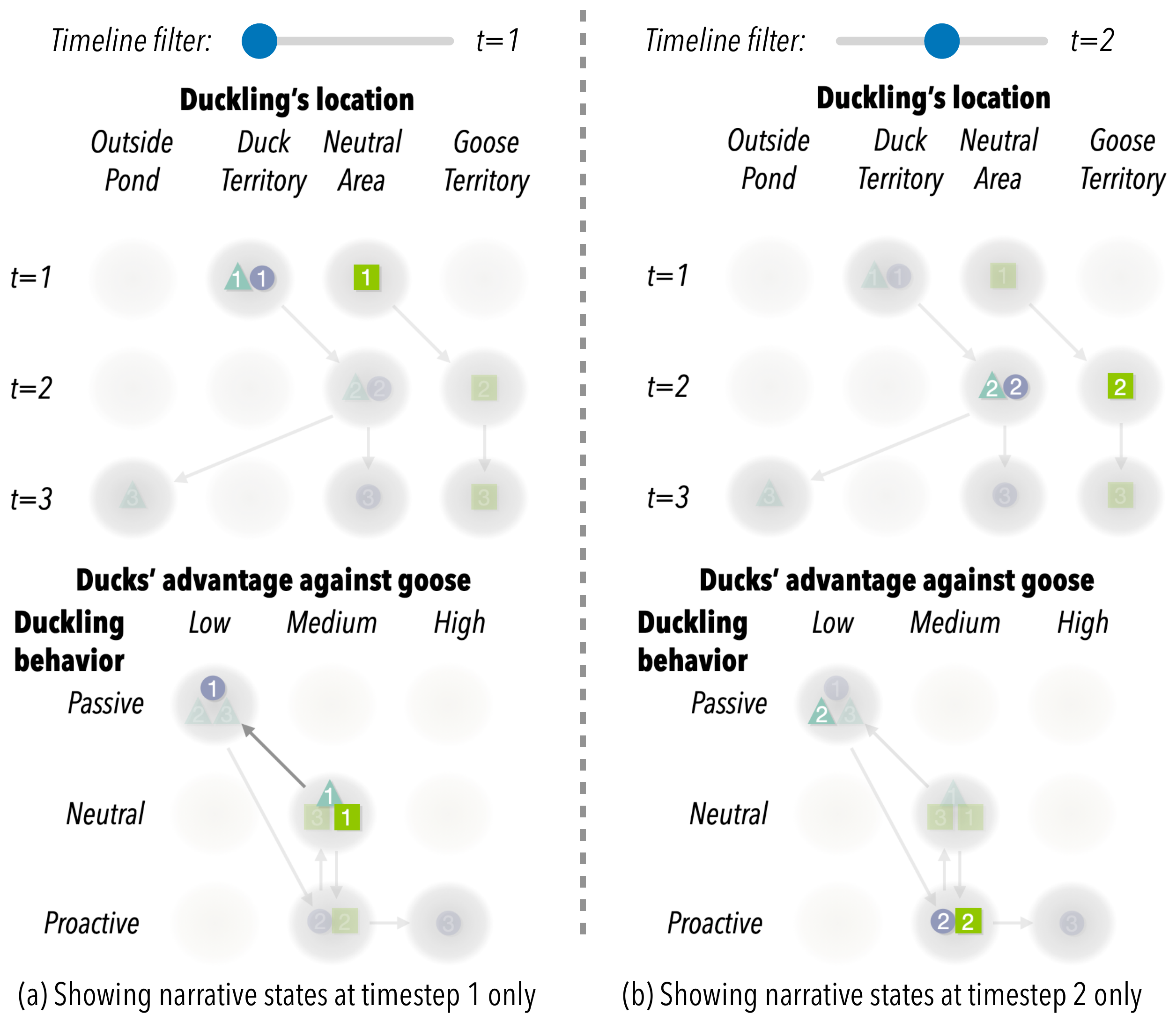}
\caption{Multiple BSV views with timeline control slider. The example illustrates how the duckling's location, the duck's competitive advantage over the goose, and the duckling's behavior co-evolve over time---revealing that high advantage against the goose is only achieved late in the story, following a period of proactive duckling behavior.}
\Description{Two side-by-side BSV diagrams, labeled (a) and (b), each controlled by a timeline filter slider shown at the top. The slider highlights narrative states at a specific timestep: t=1 in sub-figure (a) and t=2 in sub-figure (b), with states at other timesteps faded. Colored dots within gray bubbles represent narrative states from different storylines, with teal, blue, and green each corresponding to a distinct storyline. Each sub-figure has a top grid showing Duckling's Location (columns: Outside Pond, Duck Territory, Neutral Area, Goose Territory) across three timesteps (t=1, t=2, t=3), and a bottom grid showing Duckling behavior (rows: Passive, Neutral, Proactive) against Ducks' advantage against goose (columns: Low, Medium, High). Sub-figure (a) filters to timestep t=1. In the top grid, the active states at t=1 are in Duck Territory (containing teal and blue storyline dots) and Neutral Area (containing a green storyline dot). Arrows lead forward: from Duck Territory at t=1 diagonally to Neutral Area at t=2 (teal and blue dots), and from Neutral Area at t=1 diagonally to Goose Territory at t=2 (faint). From t=2, a downward arrow leads from Neutral Area to Outside Pond at t=3, and from Goose Territory to Goose Territory at t=3. In the bottom grid, the active state at Passive/Low contains a blue storyline dot. Arrows lead diagonally down-right to Neutral/Medium (green and blue dots), with a bidirectional arrow at Proactive/Medium and a rightward arrow continuing to Proactive/High. Sub-figure (b) filters to timestep t=2. In the top grid, t=1 states are faded, and the active states at t=2 are in Neutral Area (teal and blue dots) and Goose Territory (green dot). Arrows lead downward from Neutral Area at t=2 to Outside Pond at t=3, and from Goose Territory at t=2 to Goose Territory at t=3. In the bottom grid, the active state at Passive/Low contains a teal storyline dot. Arrows lead diagonally down-right to Neutral/Medium (faint), with a bidirectional arrow at Proactive/Medium (blue and green dots) and a rightward arrow to Proactive/High. Together, the two views illustrate how the duckling's location, the ducks' competitive advantage over the goose, and the duckling's behavior co-evolve: high advantage against the goose appears only at later timesteps, following a period of proactive duckling behavior.}
~\label{multiple_bsv_with_timeline_slider}
\vspace{-15pt}
\end{figure}

\begin{figure}[t]
\centering
\includegraphics[width=0.50\textwidth]{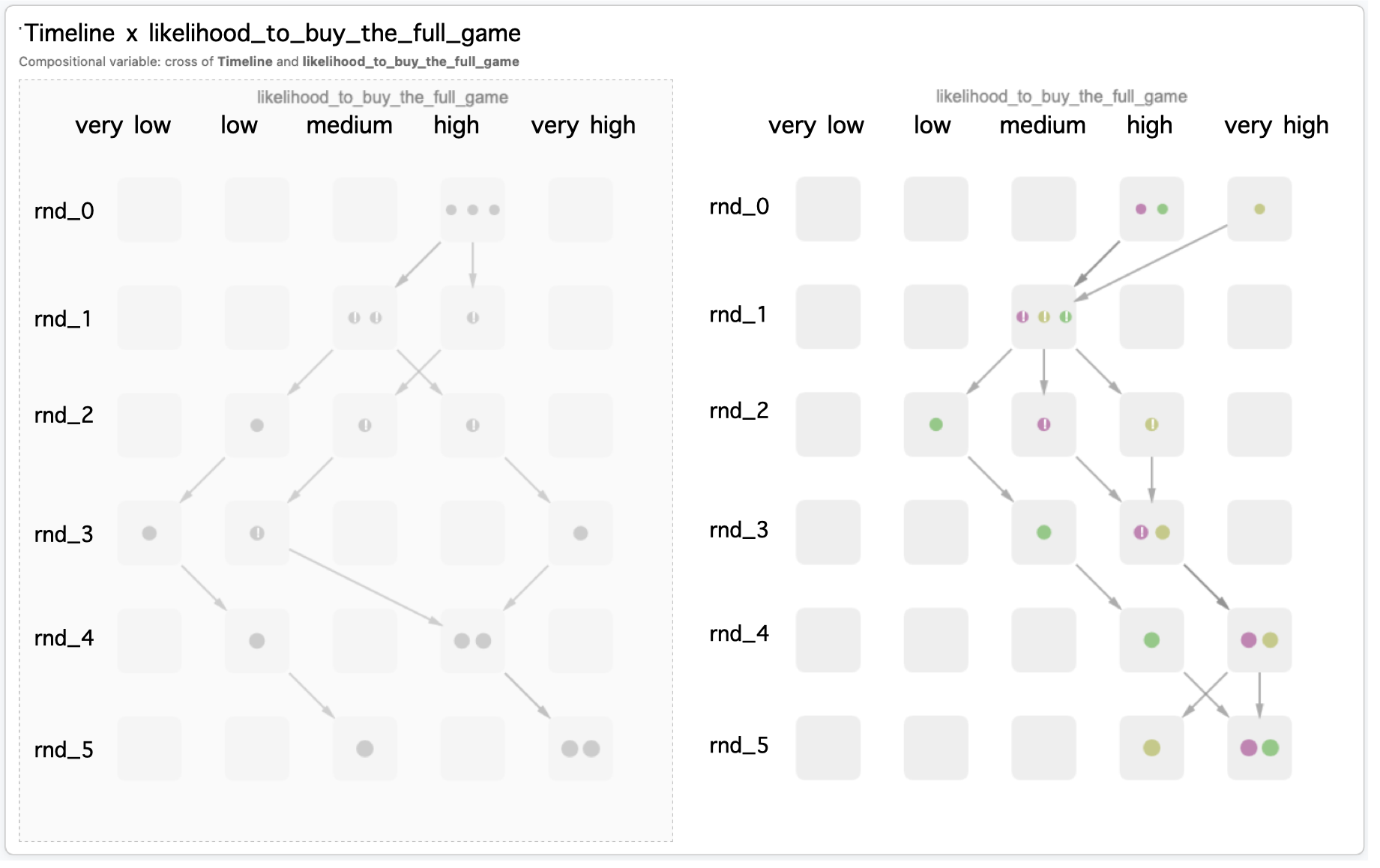}
\vspace{-10pt}
\caption{Distribution/trajectory comparison feature}
\Description{Example of distribution/trajectory comparison feature in BSV views showing two side-by-side grids. Both grids share the same axes: rows represent timeline rounds (rnd\_0 through rnd\_5) and columns represent levels of the variable likelihood\_to\_buy\_the\_full\_game (very low, low, medium, high, very high). A header label reads "Timeline x likelihood\_to\_buy\_the\_full\_game" with the subtitle "Compositional variable: cross of Timeline and likelihood\_to\_buy\_the\_full\_game." Colored dots within rounded square cells represent narrative states from different storylines; arrows between cells indicate transitions across rounds. The left grid displays all storylines rendered in gray, showing the aggregate distribution without storyline differentiation. At rnd\_0, three gray dots appear in the high column. Arrows lead to rnd\_1, where two dots appear in medium and one in high. From rnd\_1, crossing arrows lead to rnd\_2, which has dots spread across low, medium, and high. From rnd\_2, arrows fan out to rnd\_3, where dots appear in very low, low, and high. rnd\_4 shows dots in low and high, and rnd\_5 shows dots in medium and high. The right grid displays the same visualization with storylines differentiated by color: pink (magenta) and green. At rnd\_0, two pink dots appear in high and one green dot in very high. Arrows converge to rnd\_1, where three pink dots appear in medium. From rnd\_1, arrows lead to rnd\_2, which shows a green dot in very low, a half-pink half-green dot in medium, and a green dot in high. From rnd\_2, arrows lead to rnd\_3, with a green dot in medium and a half-pink half-green dot alongside another dot in high. From rnd\_3, arrows lead to rnd\_4, where a green dot appears in high and two pink dots appear in very high. From rnd\_4, crossed arrows lead to rnd\_5, where a green dot remains in high and two pink dots remain in very high. Together, the two grids illustrate how the trajectory comparison feature reveals storyline-level differences in how likelihood to buy the full game evolves over time, which are obscured in the aggregate gray view.}
~\label{trajectory_comparison}
\vspace{-20pt}
\end{figure}

Figure~\ref{multiple_bsv_with_dimension_value_filter} and Figure~\ref{multiple_bsv_with_storyline_filter} demonstrates the two narrative state highlighting mechanism described in Sec.~\ref{sec:dimension_cross_referencing}. Figure~\ref{multiple_bsv_with_timeline_slider} demonstrates the timeline control slider. Figure~\ref{trajectory_comparison} shows how each individual BSV supports narrative state distribution comparison before-and-after a new batch of playthrough data is applied.

\begin{figure}[tbh]
\centering
\includegraphics[width=0.5\textwidth]{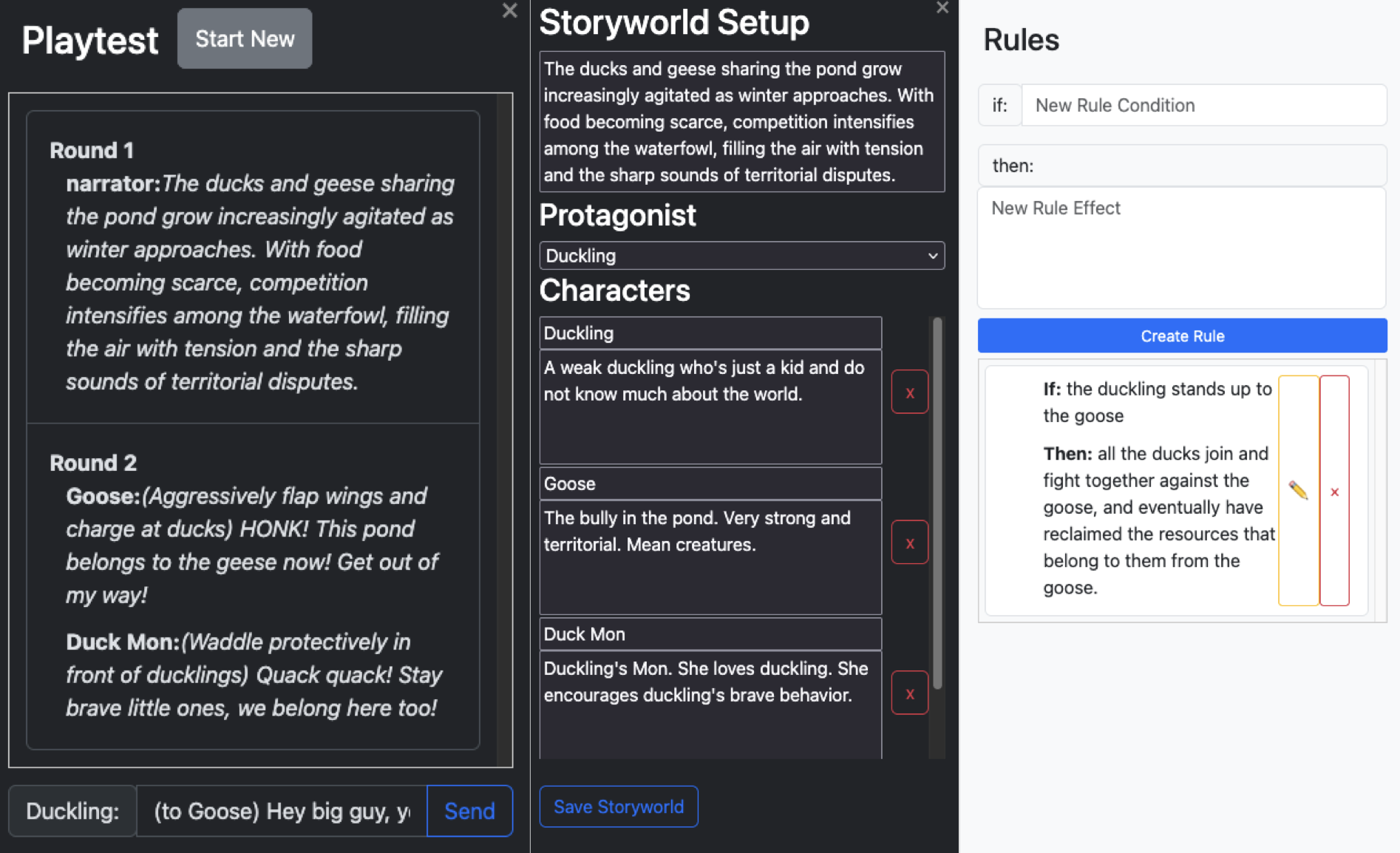}
\vspace{-10pt}
\caption{Baseline system}
\Description{This figure shows the baseline system interface for interactive narrative authoring, displaying the traditional approach without the Bundled Storyline Visualization features. The left panel contains the Playtest section with a "Start New" button and displays the unfolding narrative in a linear text format. Round 1 shows the narrator describing increasing tension at the pond as winter approaches and food becomes scarce. Round 2 presents dialogue from different characters - the Goose aggressively claiming territory and the Duck Mom protectively encouraging the ducklings. At the bottom, there's an input field where the Duckling character can enter their response "(to Goose) Hey big guy, y" with a "Send" button to submit the action. The center panel shows the Storyworld Setup interface with a dark background. It begins with the same narrative setup text about pond tensions. Below this, there's a dropdown menu for selecting the Protagonist, currently set to "Duckling." The Characters section lists three characters with their descriptions: Duckling (described as weak and unaware of the world), Goose (the pond bully who is strong and territorial), and Duck Mom (who loves and encourages the duckling's brave behavior). Each character entry has an X button for deletion. The right panel displays the Rules interface for defining narrative logic. It shows the standard if-then structure for creating rules, with buttons for "New Rule Condition" and "New Rule Effect." A "Create Rule" button appears below. An example rule is partially visible, showing "If: the duckling stands up to the goose" and "Then: all the ducks join and fight together against the goose, and eventually have reclaimed the resources that belong to them from the goose.}
~\label{baseline}
\vspace{-20pt}
\end{figure}

\subsection{Additional Information on User Study}
Table~\ref{tab:study_procedure} outlines our study procedure.
\begin{table}[tbh]
\small
% \resizebox{\linewidth}{!}{
\caption{Study Procedure}
\begin{tabular}{p{3cm} | p{1.5cm} | p{3cm} }
\hline
{\bf Activity} & {\bf Duration} & {\bf Note} \\ \hline
Set-up and Pre-study Survey & 5 min and offline time & \\ \hline
Authoring Task and Post-authoring Survey & 80 min & includes tutorial (5 min), authoring (30 min) and post-authoring survey (5 min) for each of the two conditions\\ \hline
Exit Interview & 10 min & \\ \hline
\end{tabular}
\vspace{2pt}
\Description{A three-column table titled "Study Procedure" with columns for Activity, Duration, and Note. Row 1: Set-up and Pre-study Survey, lasting 5 minutes plus offline time, with no additional notes. Row 2: Authoring Task and Post-authoring Survey, lasting 80 minutes, which includes a tutorial (5 minutes), authoring (30 minutes), and a post-authoring survey (5 minutes) for each of the two conditions. Row 3: Exit Interview, lasting 10 minutes, with no additional notes.}
\label{tab:study_procedure}
% \vspace{-25pt}
\end{table}

Table~\ref{tab:player_archetypes} lists the player profiles implemented for the user study. For some stories in our study, we had to disable the simulated killer player profile, as it frequently triggered Claude's content safety filters.
\begin{table}[hb]
\small
% \resizebox{\linewidth}{!}{
\caption{Profiles of Simulated AI Player Implemented in \system{}}
\begin{tabular}{c|p{6cm}}
{\bf Profile Name} & {\bf Description} \\\hline\hline
{\bf Role-players} & prioritize narrative immersion and character
development by mimicking the actions their character would
take in the gaming world. \\ \hline
{\bf Explorers} & motivated by curiosity and derive enjoyment from discovering, mapping, and understanding the game world, its systems, and hidden possibilities. \\\hline
{\bf Killer players}  &  motivated by competition and domination, deriving enjoyment from imposing on, defeating, or disrupting other players within the game world.\\\hline
{\bf Clueless players} & engage with a game without a clear understanding of its systems, goals, or optimal strategies, often acting through trial-and-error, guesswork, or playful misunderstanding rather than informed decision-making.
\end{tabular}
\Description{This table presents four distinct AI player profiles implemented in the Elsewise system to simulate different player behaviors. Role-players represent players who prioritize narrative immersion and character development. They engage with the story by mimicking the actions their character would realistically take within the game world. Explorers are driven by curiosity and derive satisfaction from discovering, mapping, and understanding the game world's systems and hidden possibilities. Killer players are motivated by competition and domination, finding enjoyment in imposing their will on the game world through defeating, disrupting, or otherwise asserting dominance over other players or characters within the narrative. Clueless players engage with the game without a clear understanding of its systems, goals, or optimal strategies. They navigate through the narrative using trial-and-error, guesswork, or playful misunderstanding rather than informed decision-making.}
\vspace{2pt}

\label{tab:player_archetypes}
% \vspace{-25pt}
\end{table}

Table~\ref{tab:story_themes} lists the story themes that the participants can select from in out study.
\begin{table}[b]
\small
% \resizebox{\linewidth}{!}{
\caption{Story theme candidates provided to the participants}
\begin{tabular}{p{8cm}}
Kindness is never wasted.\\\hline
Honesty always finds a way to surface.\\\hline
True friendship means accepting someone's flaws.\\\hline
Purpose gives meaning to any path.\\\hline
Kindness is not always rewarded.\\\hline
Small acts of courage can change everything.\\\hline
The journey matters more than the destination.\\\hline
True friendship means helping someone overcome their flaws.\\\hline
True courage requires accepting what cannot be changed.\\\hline
Sometimes protecting others requires withholding truth.\\\hline
\end{tabular}
\vspace{2pt}
\Description{This table describes the list of story themes provided to the participants to choose for their authoring task. The themes are: Kindness is never wasted; Honesty always finds a way to surface; True friendship means accepting someone's flaws; Purpose gives meaning to any path; Kindness is not always rewarded; Small acts of courage can change everything; The journey matters more than the destination; True friendship means helping someone overcome their flaws; True courage requires accepting what cannot be changed; Sometimes protecting others requires withholding truth.}
\label{tab:story_themes}
% \vspace{-25pt}
\end{table}

Figure~\ref{baseline} shows the baseline user interface from our user study.

\begin{figure}[t]
\centering
\includegraphics[width=0.48\textwidth]{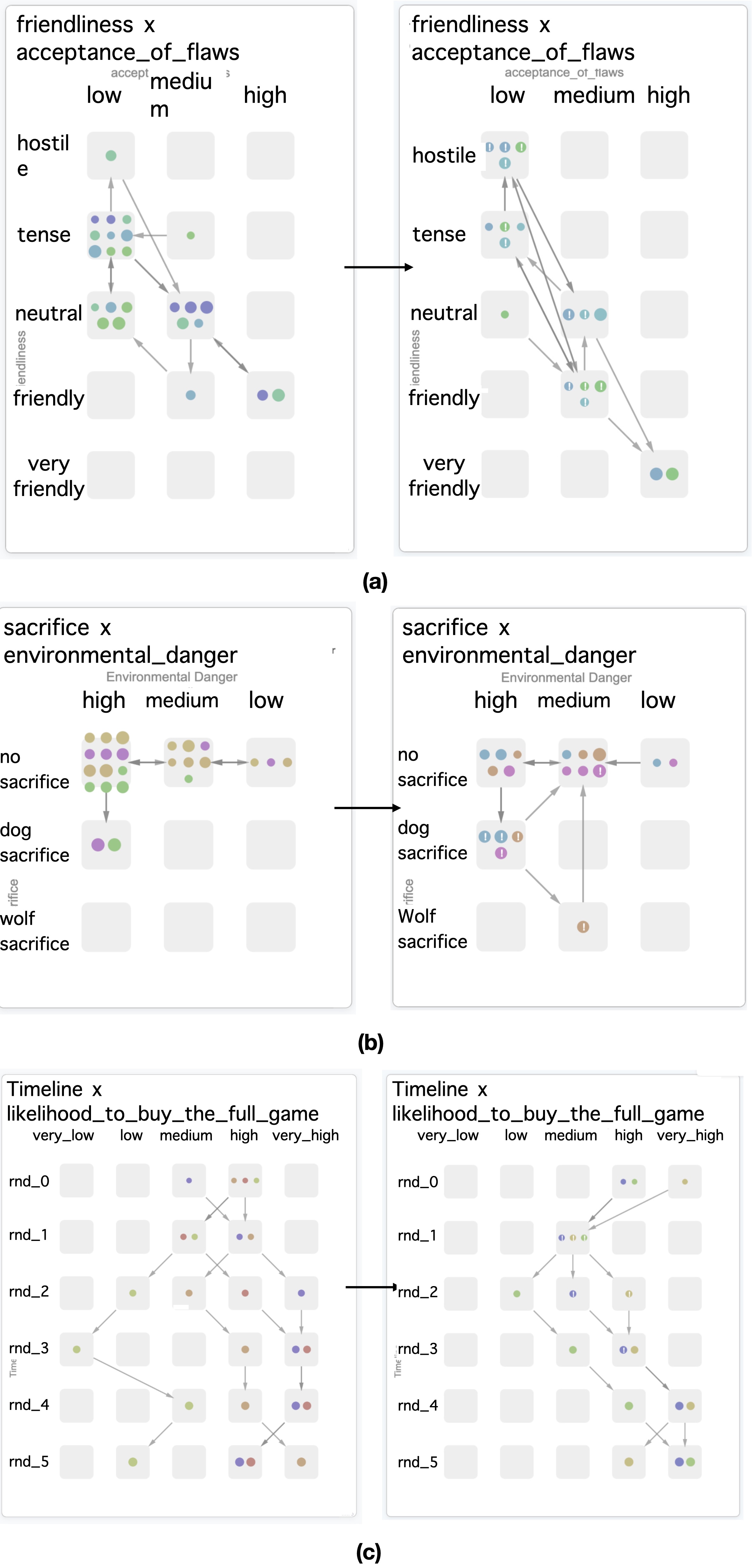}
\vspace{-10pt}
\caption{Examples of narrative state distribution before and after rule authoring. (a) P4 attempted to strengthen the positive correlation between friendliness and acceptance of flaws to communicate ``\textit{True friendship means accepting someone's flaws}''; (b) P7 tried to nudge the characters towards sacrificing for the team in dangerous situations to convey the theme ``\textit{Purpose gives meaning to any path}''; (c) P10 attemptes to guide storylines towards ``player very likely to buy the game'' at later stages on timeline.}
\Description{This figure demonstrates how the Bundled Storyline Visualization helps authors understand and refine narrative state distributions through rule authoring, showing before and after comparisons for two participants working with different thematic goals. Panel (a) shows Participant 4's attempt to strengthen the correlation between friendliness and acceptance of flaws to convey the theme "True friendship means accepting someone's flaws." The left visualization displays the initial state distribution across a 2D grid combining friendliness levels (hostile, tense, neutral, friendly, very friendly) with acceptance of flaws (low, medium, high). Colored dots represent narrative states, with connections showing possible story trajectories. The initial distribution shows scattered states without a clear positive correlation between the two dimensions. After rule authoring, shown on the right with an arrow indicating the transformation, the distribution shifts notably. More narrative states now cluster along a diagonal pattern from hostile/low acceptance to very friendly/high acceptance, demonstrating a stronger positive correlation. The connecting lines between states also show more coherent progression paths that reinforce the intended theme. Panel (b) illustrates Participant 7's efforts to convey "Purpose gives meaning to any path" by nudging characters toward team sacrifice in dangerous situations. The visualization uses sacrifice types (no sacrifice, dog sacrifice, wolf sacrifice) crossed with environmental danger levels (high, medium, low). The initial distribution on the left shows narrative states spread across various combinations without only two state being in dog sacrifice in high danger. Following rule authoring, the right visualization reveals a more intentional distribution. States now concentrate in patterns where higher environmental danger correlates with sacrifice behaviors, particularly dog sacrifice appearing more frequently in high-danger situations. The wolf sacrifice option remains less common but shows clearer contextual placement in medium danger. The refined trajectory lines between states suggest more purposeful narrative progression where dangerous situations naturally lead to sacrifice decisions.}
\label{trajectory_improvement_examples_all}
\vspace{-25pt}
\end{figure}

Figure~\ref{trajectory_improvement_examples_all} shows examples of changing narrative state distribution  resulted from participants' rule authoring.

\subsection{Demographic Information of Participants}
\begin{figure*}[tbh]
 \centering
\includegraphics[width=0.92\textwidth]{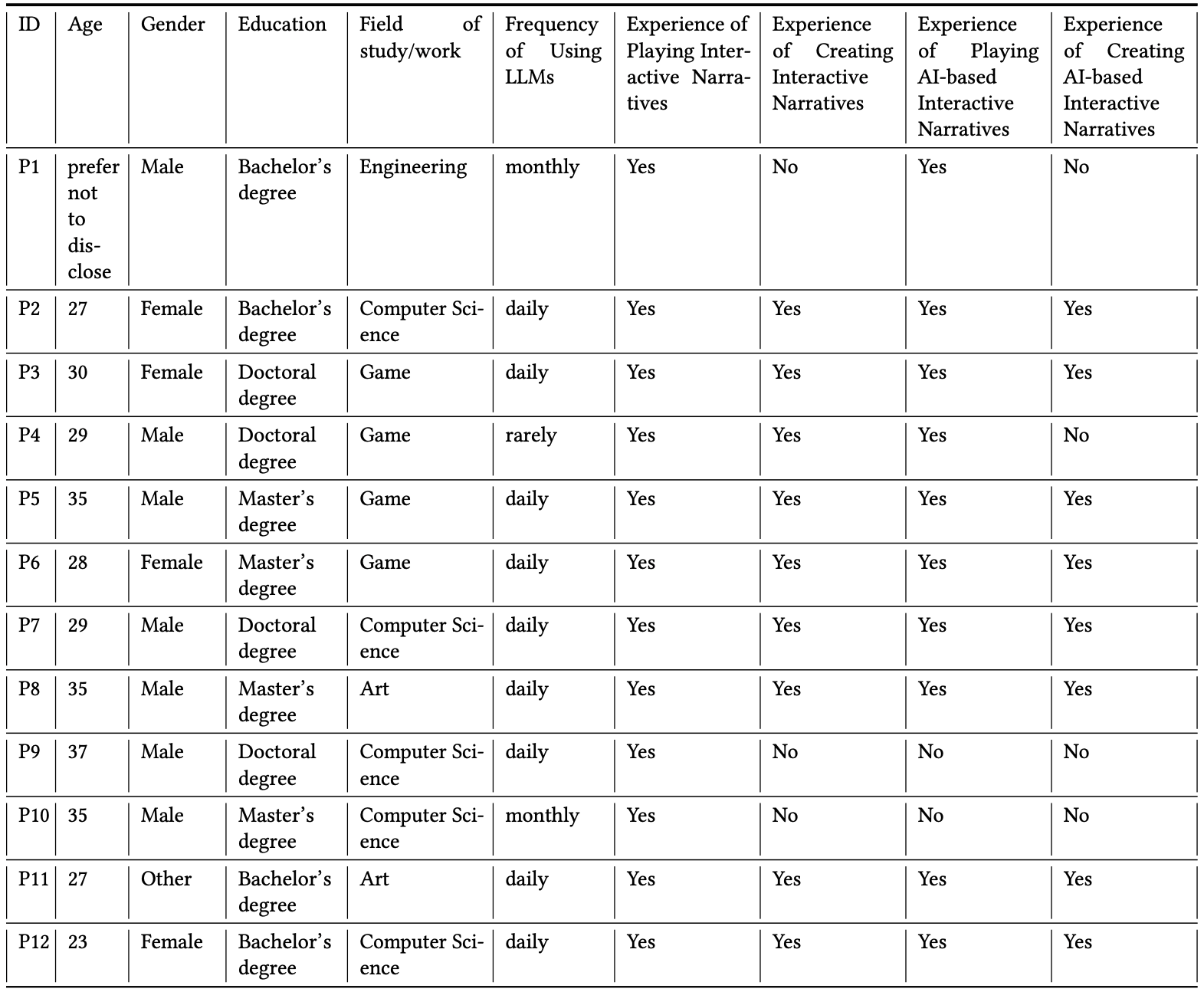}
\caption{Demographic Information of Participants}
\label{fig:participant_table}
\end{figure*}

Figure~\ref{fig:participant_table} lists detailed demographic information of each participant in our study.
% \usepackage{booktabs}
% \usepackage{longtable}
% Note: It may be necessary to compile the document several times to get a multi-page table to line up properly

\clearpage

\end{document}